\documentclass[aip, jcp, amsmath, amssymb,
 reprint
 ]{revtex4-1}

\usepackage[dvipdfmx]{graphicx}
\usepackage{dcolumn}
\usepackage{bm}

\usepackage{here}
\usepackage{color}

\begin{document}


\title[]{Phase-sensitive tip-enhanced sum frequency generation spectroscopy using temporally asymmetric pulse for detecting weak vibrational signals}

\author{Atsunori Sakurai}
\email{asakurai@ims.ac.jp}
\affiliation{Institute for Molecular Science, National Institutes of Natural Sciences, Okazaki 444-8585, Japan}
\affiliation{Graduate Institute for Advanced Studies, SOKENDAI, Okazaki 444-8585, Japan}
\affiliation{Laser-Driven Electron-Acceleration Technology Group, RIKEN SPring-8 Center, Sayocho 679-5148, Japan}
\author{Shota Takahashi}
\affiliation{Institute for Molecular Science, National Institutes of Natural Sciences, Okazaki 444-8585, Japan}
\author{Tatsuto Mochizuki}
\affiliation{Institute for Molecular Science, National Institutes of Natural Sciences, Okazaki 444-8585, Japan}
\affiliation{Graduate Institute for Advanced Studies, SOKENDAI, Okazaki 444-8585, Japan}
\author{Tomonori Hirano}
\affiliation{Department of Chemistry, Graduate School of Science, Tohoku University, Sendai 980-8578, Japan}
\author{Akihiro Morita}
\affiliation{Department of Chemistry, Graduate School of Science, Tohoku University, Sendai 980-8578, Japan}
\author{Toshiki Sugimoto}
\email{toshiki-sugimoto@ims.ac.jp}
\affiliation{Institute for Molecular Science, National Institutes of Natural Sciences, Okazaki 444-8585, Japan}
\affiliation{Graduate Institute for Advanced Studies, SOKENDAI, Okazaki 444-8585, Japan}
\affiliation{Laser-Driven Electron-Acceleration Technology Group, RIKEN SPring-8 Center, Sayocho 679-5148, Japan}

\date{\today}

\begin{abstract}
Vibrational sum frequency generation (SFG) spectroscopy is a powerful technique
for investigating molecular structures, orientations, and dynamics at surfaces.
However, its spatial resolution is fundamentally restricted to the micrometer scale by the optical diffraction limit.
Tip-enhanced SFG (TE-SFG) using a scanning tunneling microscope has been developed to overcome this limitation.
The acquired spectra exhibit characteristic dips originating from vibrational responses
located within the strong broadband non-resonant background (NRB),
which distorts and obscures the molecular signals.
By making the second pulse temporally asymmetric and introducing a controlled delay between the first and second laser pulses,
the NRB was effectively suppressed, which in turn amplified the vibrational response through interference
and facilitated the detection of weak vibrational signals.
This interference also made the technique phase-sensitive, enabling the determination of absolute molecular orientations.
Furthermore, forward- and backward-scattered signals were simultaneously detected,
conclusively confirming that the observed signals originated from tip enhancement rather than far-field contributions.
Finally, the signal enhancement factor in TE-SFG was estimated to be $6.3\times10^6-1.3\times10^7$, based on the experimental data.
This phase-sensitive TE-SFG technique overcomes the optical diffraction limit
and enables the investigation of molecular vibrations at surfaces with unprecedented detail.
\end{abstract}

\maketitle

\section{Introduction\label{Intro}}

Molecular structures and conformations at surfaces largely govern the physicochemical properties,
functionalities, and reactivity of materials.\cite{Somorjai_book}
Therefore, the development of advanced methodologies capable of resolving molecular-level details
at surfaces is indispensable for elucidating the microscopic mechanisms underlying surface phenomena.
Vibrational sum frequency generation (SFG) spectroscopy is a powerful technique
for characterizing molecular structures, orientations,
and ultrafast dynamics at surfaces and interfaces.\cite{Shen_nature89, Shen_SFG_book, Morita_book}
The inherent surface specificity of this method arises from the second-order nonlinear susceptibility, $\chi^{(2)}$,
which vanishes in centrosymmetric media under the dipole approximation.\cite{Shen_nature89, Shen_SFG_book, Morita_book}
Moreover, SFG is applicable to time-resolved spectroscopy\cite{Guyot-Sionnest_PRL90, Ueba_PSS97, Bonn_PRL00, Inoue_PRL16}
and has the unique ability to enable the determination of absolute (up \textit{vs.} down)
molecular orientations.\cite{Harris_CPL87, Shen_PRB99, Shen_JACS07, Tahara_JCP09, Sugimoto_NatPhys16, Yamaguchi_PCCP21}
Owing to its versatility, SFG has been extensively employed to investigate a wide range of systems,
as comprehensively discussed in previous reviews.
\cite{Shen_ARPC02, Shen_CR06, Geiger_ARPC, Tahara_ARPC13, Weidner_CR20, Sugimoto_PCCP20, Borguet_JPCC22}

Recently, considerable efforts have been focused on combining SFG with optical microscopy to probe structural heterogeneities
at the microscopic level.\cite{Baldelli_JPCC09, Potma_OL11, Ge_JPCB21, Wang_ARPC21, Thamer_OE23}
As a distinct nonlinear optical modality,
SFG has been employed for multimodal bioimaging in combination with coherent anti-Stokes Raman scattering (CARS)
and two-photon fluorescence (2PFL) techniques.\cite{Herdzik_APB20}
Furthermore, phase-resolved SFG microscopy has revealed the three-dimensional architecture of microscale membrane domains,
characterized by orientational order associated with molecular chirality.\cite{Alex_NatComm}
Despite these remarkable advances,
the spatial resolution attainable with conventional far-field optical spectroscopies is fundamentally restricted
by the optical diffraction limit.
Consequently, the extension of SFG to nanoscale spectroscopy for resolving structural inhomogeneities
below the optical wavelength remains a challenge.

To overcome the diffraction limit,
scanning near-field optical microscopy (SNOM) has emerged as an advanced optical imaging technique
providing nanoscale spatial resolution.\cite{Richards_book08, Atkin_AdvPhys12}
SNOM integrates optical microscopy with scanning probe techniques,
including scanning tunneling microscopy (STM) and atomic force microscopy (AFM).
Particularly, when light scattered from a sharp metallic tip is employed for detection,
the resolution of this scattering-type SNOM (\textit{s}-SNOM) is wavelength-independent
and fundamentally limited by the radius of curvature of the tip apex.\cite{Hillenbrand_NL06}
\textit{s}-SNOM is often referred to as tip-enhanced spectroscopy
when the plasmonic field enhancement plays a pivotal role in exciting local optical processes
with different excitation and emission wavelengths.\cite{Atkin_AdvPhys12, Richards_book08}
Both \textit{s}-SNOM
\cite{Atkin_AdvPhys12, Richards_book08, Keilmann_nature99, Basov_science07, Hillenbrand_nature12, Basov_nature12, Basov_AdvMat19}
and tip-enhanced spectroscopy
\cite{Richards_book08, Hartschuh_Angew08, Weckhuysen_NatNano12, Deckert_CSR17, Verma_CR17, Park_nanophoton20}
have been broadly applied in chemical imaging and to investigate the nanoscale optical properties of materials.
In particular, when the substrate is metallic,
the nanogap formed between it and the metallic tip acts as a cavity that confines the electric field,
producing a strong field enhancement through gap-mode plasmon resonance.\cite{Li_AdvMat24}
This resonance is crucial for achieving ultrahigh-resolution optical imaging,
as evidenced by single-molecule visualization via Raman\cite{Hou_nature13, Apkarian_nature19, Rafael_NatNano20}
and fluorescence\cite{Hou_NatPhoton20} detection.
Furthermore, optical nanoimaging based on multimodal nonlinear optical processes
(second-harmonic generation (SHG), SFG, four-wave mixing, and two-photon photoluminescence)
has been actively explored by Wang, El-Khoury, and their coworkers.\cite{El-Khoury_JPCL21, El-Khoury_JPCL21_2, El-Khoury_JPCL22, El-Khoury_JPCC23}

However, gap-mode plasmons typically appear in the visible and near-IR regions\cite{Imada_PRL17, Kumagai_PRL18}
and do not effectively enhance the IR signals.
This spectral limitation poses a challenge for IR vibrational spectroscopy,
as molecular vibrations are located in the IR region.
To detect weak near-field signals,
IR \textit{s}-SNOM employs an interferometric technique\cite{Hillenbrand_APL06, Dai_ApplSpecRev18}
and lock-in detection.\cite{Dai_ApplSpecRev18, Nishida_NL24}

The SFG process represents a promising strategy for enhancing IR signals,
because it upconverts vibrational responses from the IR to the visible region,
where the gap-mode plasmons are active.\cite{Takahashi_JPCL23, Sakurai_NL25}
A related concept has been proposed for molecules embedded in an optomechanical nanocavity,
which acts as a coherent frequency converter,
where molecular vibrations serve as mechanical oscillators.\cite{Galland_science21, Baumberg_science21, Oyamada_arXiv}
In a previous study,\cite{Takahashi_JPCL23}
we investigated tip-enhanced SHG spectroscopy across a broad spectral range
and identified two enhancement mechanisms underlying tip-enhanced spectroscopy:
(i) the incident light is amplified when received by the tapered shaft of the tip and guided to the apex (antenna effect);
(ii) the radiation efficiency is enhanced by the gap-mode plasmon (plasmonic resonance).
Numerical simulations revealed that the antenna effect becomes more pronounced at longer wavelengths,
indicating that the IR region provides significant enhancement of the incident field,
whereas plasmonic resonance improves the radiation efficiency in the visible region.\cite{Takahashi_JPCL23}
Nevertheless, the use of tip-enhanced SFG (TE-SFG) for detecting molecular vibrations remains relatively unexplored.
To the best of our knowledge, Fellows first reported the application of TE-SFG to molecules in 2022;\cite{Alex_PhD}
however, in that study the use of plasmonic resonance was limited because the sample molecules were deposited on the CaF$_2$ substrate.

Thereafter, we implemented TE-SFG and detected signals originating
from molecules located within the STM nanojunction formed between a gold substrate and a gold tip.\cite{Sakurai_NL25}
The acquired signal consisted of two contributions:
vibrationally resonant responses from the molecules and vibrationally non-resonant responses from the gold substrate and/or tip.
The interference between these two components produced a molecular vibrational signature that appeared as a dip
within a broad SFG signal arising from the gold-derived non-resonant background (NRB).
For metallic surfaces, this NRB often exhibits high intensity due to surface plasmon resonance.
The NRB is widely recognized as a major limitation in SFG spectroscopy,
as it distorts and obscures the molecular vibrational features of interest.
\cite{Harris_CPL87, Patterson_JPCC11_1, Patterson_JPCC11_2, Benderskii_JPCL12, Patterson_JCP24}
Therefore, an NRB suppression technique using temporally asymmetric pulses has been developed in far-field SFG measurements.\cite{Dlott_JPCC07}
Conversely, other studies have shown that, under certain conditions,
the NRB can be exploited to amplify weak vibrational signals through interferometric enhancement.\cite{Harris_CPL87, Patterson_JPCC11_2}
Furthermore, if the relative NRB phase is known,
this approach enables phase-sensitive detection of the real and imaginary components of $\chi^{(2)}$,
thereby allowing the determination of absolute molecular orientations.
\cite{Harris_CPL87, Castner_Langmuir10, Patterson_JPCC11_1, Patterson_JPCC11_2, Benderskii_JPCL12}

In this study, we demonstrated the applicability of an interferometric approach to TE-SFG
by introducing a controlled time delay between the first (IR) and second (temporally asymmetric) pulses.
This method enabled the detection of a weak vibrational mode,
previously unresolved,\cite{Sakurai_NL25}
whose associated dip feature became increasingly pronounced at longer time delays.
Owing to the much smaller number of molecules (likely fewer than one hundred)
located within the STM nanojunction compared to typical far-field measurements ($\sim$10$^6$),
the corresponding molecular signal is inherently weak.
Moreover, the use of the gap-mode plasmon for signal enhancement is unavoidably accompanied by a substantial NRB.
Therefore, effective suppression of the NRB combined with interferometric amplification of the molecular response
is crucial for achieving both high sensitivity and nanoscale spatial resolution in TE-SFG.

This paper is organized as follows: In Sec.~\ref{sec2},
we provide a brief outline of the SFG process in the time and frequency domains,
the mechanisms for NRB suppression and signal amplification using a temporally asymmetric pulse,
and the signal enhancement mechanism in TE-SFG.
Sec.~\ref{sec3} describes the experimental setup.
In Sec.~\ref{sec4}, we present and discuss the experimental results
on the spectral acquisition, IR–visible pulse delay effect,
molecular orientation, simultaneous detection of forward- and backward-scattered signals, and signal enhancement factor.
Finally, Sec.~\ref{sec5} presents the concluding remarks.

\section{Theoretical Background\label{sec2}}
\subsection{General description of vibrational SFG process in time and frequency domains}

\begin{figure}
 \includegraphics[width=6cm]{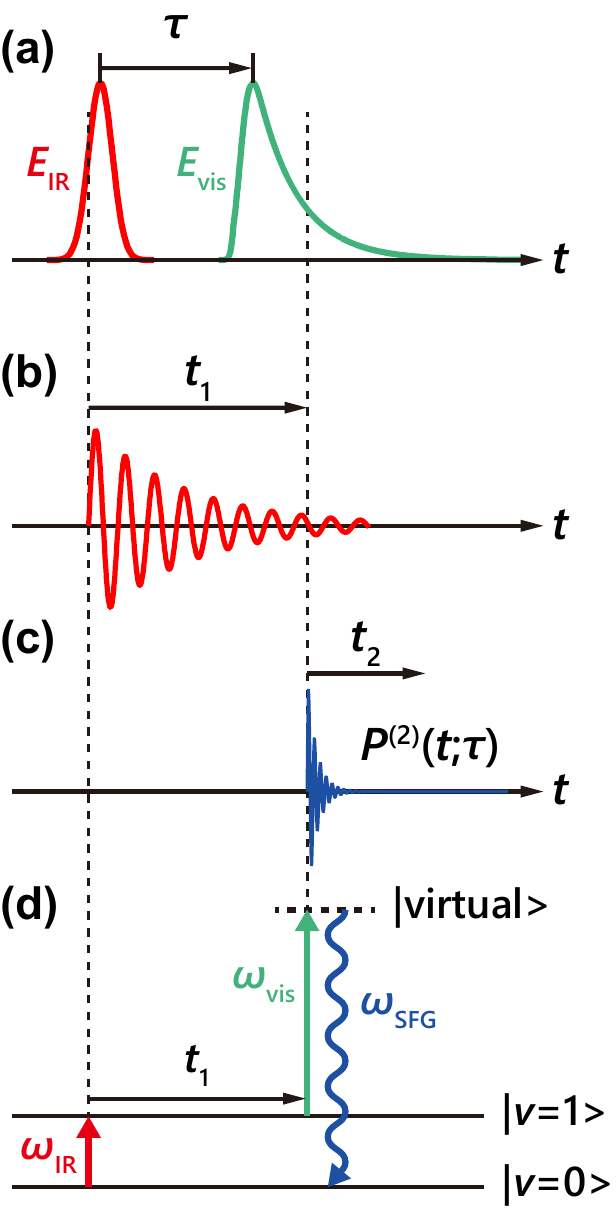}
 \caption{(a) IR and visible laser fields, with a time delay between the peaks of the two pulses.
 (b) Vibrational polarization generated by excitation with the IR pulse.
 (c) Second-order polarization induced by interaction with the visible pulse.
 (d) Energy level diagram illustrating the SFG process.
 The dashed line corresponds to the virtual state associated with anti-Stokes Raman scattering.}
 \label{fig1}
\end{figure}

Vibrational SFG is a second-order nonlinear spectroscopic technique
in which molecular vibrations are resonantly excited by an IR pulse
and coherently upconverted to the visible signal
by a second pulse through anti-Stokes Raman scattering.
The second pulse is typically in the visible or near-IR range,
but has historically been referred to as ``visible,'' and here we follow this convention.
The polarization generated by the SFG process is expressed as\cite{Mukamel}
\begin{align}
 P^{(2)}(t;\tau)
  = \epsilon_0 & \int_0^\infty dt_2 \int_0^\infty dt_1\, \chi^{(2)}(t_2,t_1)  \notag \\
 &\times E_\mathrm{vis}(t-t_2;\tau)E_\mathrm{IR}(t-t_2-t_1),  \label{P2}
\end{align}
where $P^{(2)}(t;\tau)$ is the second-order polarization, $\epsilon_0$ is the permittivity of vacuum,
and $\chi^{(2)}(t_2,t_1)$ is the second-order response function of the system in the time domain,
while $E_\mathrm{vis}(t)$ and $E_\mathrm{IR}(t)$ are the visible and IR laser fields, respectively.
Moreover, the time $t_1$ represents the interval between the first interaction with the IR pulse and the second interaction with the visible pulse,
during which the excited vibrational polarization persists (Fig.~\ref{fig1}(b)).
The subsequent time $t_2$ denotes the period from the second interaction to the detection time $t$ (Fig.~\ref{fig1}(c)).
Herein, we define the time delay $\tau$ as the interval between the peaks of the first IR and second visible pulses,
which is experimentally tunable (Fig.~\ref{fig1}(a)).
Broadband SFG measurements employ a spectrally broadband IR pulse and a narrowband visible pulse,\cite{Richter_OL98, Wang_PCCP13}
so that the visible pulse has a longer duration than the IR one.
The time profiles of the two laser pulses,
the excited vibrational polarization,
the second-order polarization $P^{(2)}(t;\tau)$,
and the corresponding energy level diagram\cite{Benderskii_JPCB05} are shown in Fig.~\ref{fig1}.

In the SFG process, the second interaction with the visible pulse is off-resonant
with electronic transitions (i.e., transitions to virtual states);
therefore, the electronic response decays so quickly that, to a good approximation, it can be treated as instantaneous.
Under this assumption, the system response can be approximated by\cite{Wang_PCCP13, Benderskii_JPCB05, Benderskii_JCP10, Zanni_JPCB11}
\begin{equation}
 \chi^{(2)}(t_2,t_1) \approx 2\delta(t_2)\chi^{(2)}(t_1),
\end{equation}
where $\chi^{(2)}(t_1) = \int_0^\infty dt_2\, \chi^{(2)}(t_2,t_1)$.
In this situation, Eq.~(\ref{P2}) becomes
\begin{align}
 P^{(2)}(t;\tau)
 &= \epsilon_0 E_\mathrm{vis}(t;\tau)\int_0^\infty dt_1\, \chi^{(2)}(t_1)E_\mathrm{IR}(t-t_1)  \notag \\
 &= \epsilon_0 E_\mathrm{vis}(t;\tau)\left[ \chi^{(2)}(t)*E_\mathrm{IR}(t)\right],  \label{P2_2}
\end{align}
where the symbol $*$ denotes convolution.
Given that $E_\mathrm{SFG}(t;\tau)\propto P^{(2)}(t;\tau)$
and the SFG spectra are measured in the frequency domain with a spectrograph,
which yields the Fourier-transformed field $\tilde{E}_\mathrm{SFG}(\omega;\tau)$, it follows that
\begin{align}
 \tilde{E}_\mathrm{SFG}(\omega;\tau)
  &\propto \int_{-\infty}^\infty dt\, P^{(2)}(t;\tau)e^{i\omega t}  \notag \\
  &= \frac{\epsilon_0}{2\pi} \tilde{E}_\mathrm{vis}(\omega;\tau)*\left[\tilde{\chi}^{(2)}(\omega)\tilde{E}_\mathrm{IR}(\omega) \right]  \notag \\
  &= \frac{\epsilon_0}{2\pi}\int_{-\infty}^\infty d\omega_\mathrm{IR}\, \tilde{E}_\mathrm{vis}(\omega_\mathrm{vis};\tau)\tilde{\chi}^{(2)}(\omega_\mathrm{IR})\tilde{E}_\mathrm{IR}(\omega_\mathrm{IR}),  \label{E_SFG}
\end{align}
which is recorded with a multiplex charge-coupled device (CCD) detector, and $\omega_\mathrm{vis}=\omega - \omega_\mathrm{IR}$.
Eq.~(\ref{E_SFG}) shows that the SFG signal contains the product of the spectral response
$\tilde{\chi}^{(2)}(\omega_\mathrm{IR})$ and the IR pulse spectrum $\tilde{E}_\mathrm{IR}(\omega_\mathrm{IR})$,
which is convolved with the visible pulse spectrum $\tilde{E}_\mathrm{vis}(\omega_\mathrm{vis};\tau)$.
This means that the spectral range covered by the measurement is determined by the IR pulse spectrum $\tilde{E}_\mathrm{IR}(\omega_\mathrm{IR})$,
while the spectral resolution is determined by the spectral width of $\tilde{E}_\mathrm{vis}(\omega_\mathrm{vis})$.\cite{Benderskii_JCP10}

The second-order response function consists of two contributions:
vibrationally resonant ($\chi^{(2)}_\mathrm{R}(t)$) and non-resonant ($\chi^{(2)}_\mathrm{NR}(t)$) responses.
The resonant term is expressed as a sum of vibrational modes, each defined by its amplitude $A_q$,
resonant frequency $\omega_q$, and damping constant $\Gamma_q$ for the $q$-th mode,
while the non-resonant term is characterized by a constant amplitude $A_\mathrm{NR}$ and a phase factor $\theta_\mathrm{NR}$.
The second-order response function is expressed as\cite{Mukamel, Wang_PCCP13, Benderskii_JPCB05}
\begin{align}
 \chi^{(2)}(t)
 &=\, \chi^{(2)}_\mathrm{NR}(t) + \chi^{(2)}_\mathrm{R}(t)  \notag \\
 &=\, 2A_\mathrm{NR}e^{i\theta_\mathrm{NR}}\delta(t)
 + i\sum_q A_q e^{(-i\omega_q - \Gamma_q)t},  \label{chi1_time}
\end{align}
where $\delta(t)$ is the Dirac delta function representing the instantaneous non-resonant response.
The Fourier transform of Eq.~(\ref{chi1_time}) is expressed as
\begin{align}
 \tilde{\chi}^{(2)}(\omega)
  &= \int_0^\infty dt\, \chi^{(2)}(t)e^{i\omega t}  \notag \\
  &= A_\mathrm{NR}e^{i\theta_\mathrm{NR}}
  + \sum_q \frac{A_q}{\omega_q - \omega - i\Gamma_q}.  \label{chi1_omega}
\end{align}
Under our experimental conditions, the IR pulse has a broad bandwidth,
and $|\tilde{E}_\mathrm{IR}(\omega_\mathrm{IR})|^2$ can be approximated by a Gaussian function
with a central frequency $\omega_\mathrm{IR}^0$ and a spectral bandwidth $\sigma_\mathrm{IR}$:
\begin{equation}
 |\tilde{E}_\mathrm{IR}(\omega_\mathrm{IR})|^2
  \propto \exp\left[-\frac{(\omega_\mathrm{IR}-\omega_\mathrm{IR}^0)^2}{\sigma_\mathrm{IR}^2}\right].  \label{Gaussian}
\end{equation}
In addition, the visible pulse has a narrow bandwidth and can thus be approximated
as $\tilde{E}_\mathrm{vis}(\omega_\mathrm{vis};\tau)\propto\delta(\omega_\mathrm{vis}-\omega_\mathrm{vis}^0)$.
Under these assumptions, the SFG spectrum measured in the frequency domain can be expressed as
\begin{align}
 &I_\mathrm{SFG}(\omega)
 \propto \left|\tilde{E}_\mathrm{SFG}(\omega)\right|^2
 \propto \left|\tilde{\chi}^{(2)}(\omega_\mathrm{IR})\tilde{E}_\mathrm{IR}(\omega_\mathrm{IR}) \right|^2 \notag \\
 &\propto \left|A_\mathrm{NR}e^{i\theta_\mathrm{NR}} + \sum_q\frac{A_q}{\omega_q-\omega_\mathrm{IR}-i\Gamma_q} \right|^2
 \exp\left[-\frac{(\omega_\mathrm{IR} - \omega_\mathrm{IR}^0)^2}{\sigma_\mathrm{IR}^2}\right],  \label{I_SFG}
\end{align}
where $\omega_\mathrm{IR} = \omega - \omega_\mathrm{vis}^0$.

\subsection{NRB suppression and signal amplification using temporally asymmetric pulse}

The SFG spectra directly reflect resonant vibrational responses at surfaces and interfaces.
However, these signals are often distorted and obscured by the presence of an NRB,
which originates from the instantaneous electronic response of materials to the electric field
of the incident IR pulse.\cite{Patterson_JCP24}
The NRB arises only when the first IR and the second visible pulses temporally overlap,
whereas the resonant response persists over the vibrational decay time,
which is typically longer than the IR pulse duration.
Therefore, introducing a time delay between the two pulses can suppress the NRB.
The same approach has also been applied to suppress the NRB in CARS.\cite{Scully_Science07, Cho_JPCB23}
The efficiency of NRB suppression is influenced by the temporal profile of the visible pulse.
Additionally, the relative contrast of the vibrationally resonant signals to the NRB
can be optimized by tailoring the visible pulse shape.
Although the visible pulse was treated as a delta function in the frequency domain
(i.e., a continuous wave (CW) in the time domain) in the derivation of Eq.~(\ref{I_SFG}),
here we explicitly consider the effect of its temporal profile.

We assume that $\chi^{(2)}_\mathrm{R}(t)$ in Eq.~(\ref{chi1_time}) consists of a single vibrational mode,
and that $P^{(2)}(t;\tau)$ comprises resonant ($P^{(2)}_\mathrm{R}(t;\tau)$) and non-resonant ($P^{(2)}_\mathrm{NR}(t;\tau)$) contributions.
Then, Eq.~(\ref{P2_2}) can be written as
\begin{align}
 P^{(2)}&(t;\tau)
 = P^{(2)}_\mathrm{NR}(t;\tau) + P^{(2)}_\mathrm{R}(t;\tau)  \notag \\
 =& \, \epsilon_0 A_\mathrm{NR}e^{i\theta_\mathrm{NR}}E_\mathrm{vis}(t;\tau)E_\mathrm{IR}(t)  \notag \\
 &+ i \epsilon_0 A_q E_\mathrm{vis}(t;\tau) \int_0^\infty dt_1\, e^{(-i\omega_q - \Gamma_q)t_1} E_\mathrm{IR}(t-t_1).  \label{P2_3}
\end{align}
The first term on the right-hand side of Eq.~(\ref{P2_3}) indicates that
$P^{(2)}_\mathrm{NR}(t;\tau)$ arises from the $E_\mathrm{vis}(t;\tau)E_\mathrm{IR}(t)$ product.
The temporal profiles of the two pulses are assumed to be Gaussian, as described by the following equations:
\begin{gather}
 E_\mathrm{IR}(t)
 = E_\mathrm{IR}^0 \exp\left[-\frac{\sigma_\mathrm{IR}^2}{2} t^2\right]
 \exp\left(-i\omega_\mathrm{IR}^0t\right),  \label{E_IR} \\
 E_\mathrm{vis}(t;\tau)
 = E_\mathrm{vis}^0 \exp\left[-\frac{\sigma_\mathrm{vis}^2}{2} (t - \tau)^2\right]
 \exp\left(-i\omega_\mathrm{vis}^0(t-\tau) \right),  \label{E_vis}
\end{gather}
where $E_\mathrm{IR}^0$ and $E_\mathrm{vis}^0$ are the amplitudes of the IR and visible pulses, respectively,
while $\sigma_\mathrm{IR}$ and $\sigma_\mathrm{vis}$ are the corresponding spectral widths,
and the visible pulse is delayed by $\tau$ relative to the IR pulse.
The Fourier transform of $P^{(2)}_\mathrm{NR}(t;\tau)$ is given by
\begin{align}
 \tilde{P}^{(2)}_\mathrm{NR}(\omega;\tau)
 &\propto \int_{-\infty}^\infty dt\, E_\mathrm{vis}(t;\tau)E_\mathrm{IR}(t) e^{i\omega t}  \notag \\
 &\propto \exp\left[ -\frac{\sigma_\mathrm{vis}^2\sigma_\mathrm{IR}^2}{2(\sigma_\mathrm{vis}^2 + \sigma_\mathrm{IR}^2)}\tau^2\right]
 \exp(i\omega_\mathrm{vis}^0 \tau),  \label{P_NR}
\end{align}
where we applied the relation $\omega-\omega_\mathrm{vis}^0-\omega_\mathrm{IR}^0=0$.

\begin{figure}
 \includegraphics[width=7cm]{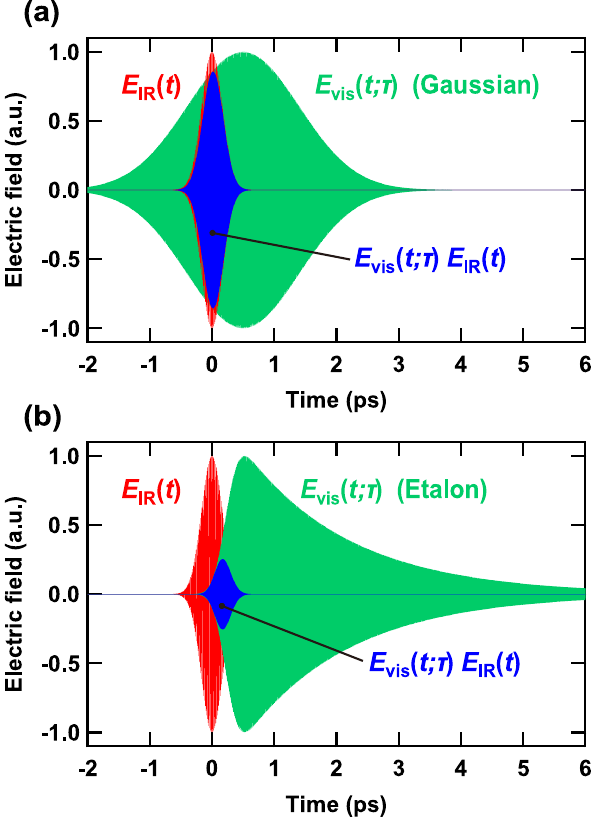}
 \caption{Temporal profiles of IR ($E_\mathrm{IR}(t)$) and visible ($E_\mathrm{vis}(t;\tau)$) pulses,
 along with their product ($E_\mathrm{vis}(t;\tau)E_\mathrm{IR}(t)$).
 Their central frequencies are $\omega_\mathrm{IR}^0=2900\, \mathrm{cm}^{-1}$
 and $\omega_\mathrm{vis}^0=9680.5\, \mathrm{cm}^{-1}$ (1033 nm).
 $E_\mathrm{IR}^0$ in Eq.~(\ref{E_IR}) and $E_\mathrm{vis}^0$ in Eq.~(\ref{etalon}) were determined
 so that the maximum amplitude of $E_\mathrm{IR}(t)$ and $E_\mathrm{vis}^\mathrm{etalon}(t;\tau)$ was equal to 1, respectively.
 (a) The IR and visible pulses are assumed to be Gaussian, with full widths at half-maximum (FWHMs) of 0.28 and 1.5 ps, respectively,
 and a time delay $\tau$ = 0.5 ps.
 (b) The visible pulse is expressed by Eq.~(\ref{etalon}),
 with an original input pulse FWHM of 0.28 ps, a central wavelength of 1033 nm,
 a reflectivity $R=0.93$, and an air gap $d=18.6\,\mu\mathrm{m}$.
 FWHMs are defined for $|E(t)|^2$, while the displayed profiles correspond to $E(t)$.}
 \label{fig2}
\end{figure}

As broadband SFG measurements use spectrally broadband IR and narrowband visible pulses
($\sigma_\mathrm{IR} \gg \sigma_\mathrm{vis}$),\cite{Richter_OL98, Wang_PCCP13}
the intensity decay of the non-resonant second-order polarization can be approximated as
$\left|\tilde{P}^{(2)}_\mathrm{NR}(\omega;\tau)\right|^2 \propto e^{-\sigma_\mathrm{vis}^2\tau^2}$.
This indicates that the NRB decay is determined by the visible pulse duration, as schematically shown in Fig.~\ref{fig2}(a).
As shown in Eq.~(\ref{E_SFG}), obtaining high-resolution SFG spectra requires a narrow spectral width of the visible pulse;\cite{Wang_PCCP13}
however, this results in a longer NRB decay.

Temporally asymmetric visible pulses created by a Fabry–P\'{e}rot etalon have proven useful to avoid this problem.\cite{Dlott_JPCC07}
The visible pulse transmitted through the etalon has a temporal profile consisting of a series of replicas of the original input pulse,
separated by the round-trip time $\tau_\mathrm{RT}=2d/c$, where $d$ is the air gap between the mirrors
and $c$ is the speed of light:\cite{Benderskii_JCP10}
\begin{align}
 E_\mathrm{vis}^\mathrm{etalon}& (t;\tau)
  = E_\mathrm{vis}^0(1-R)\sum_{n=0}^\infty R^n  \notag \\
 \times& \exp\left[-\frac{\sigma_\mathrm{vis}^2}{2}(t-n\tau_\mathrm{RT}-\tau)^2 - i\omega_\mathrm{vis}^0(t-n\tau_\mathrm{RT}-\tau) \right].  \label{etalon}
\end{align}
In Eq.~(\ref{etalon}), $R$ denotes the reflectivity of the etalon mirrors.
When the original visible pulse has the same duration as the IR pulse (i.e., $\sigma_\mathrm{vis}=\sigma_\mathrm{IR}$),
the leading edge of $|E_\mathrm{vis}^\mathrm{etalon}(t;\tau)|$ closely resembles that of the IR pulse (Fig.~\ref{fig2}(b)),
because it corresponds to the first pulse transmitted through the etalon (i.e., $n=0$ in Eq.~(\ref{etalon})).
In this case, the decay of the non-resonant second-order polarization can be approximated as
$\left|\tilde{P}^{(2)}_\mathrm{NR}(\omega;\tau)\right|^2\propto e^{-\sigma_\mathrm{IR}^2\tau^2/2}$,
which is much faster than that expected for a Gaussian visible pulse.

In addition to suppressing the NRB, interferometric detection has been shown to enhance the sensitivity of SFG spectroscopy,
as it amplifies weak responses by mixing them with a reference signal (the so-called local oscillator).\cite{Benderskii_JACS08}
This approach also makes it possible to obtain both amplitude and phase information of SFG signals,\cite{Shen_JACS07, Yamaguchi_PCCP21}
thereby enabling the determination of the absolute (up \textit{vs.} down) molecular orientation.\cite{Tahara_JCP09, Sugimoto_NatPhys16, Yamaguchi_JCP08}
Instead of using a local oscillator generated separately from the sample,
the NRB itself can serve as an internal reference, provided that its phase is known.

To illustrate this point, we numerically calculated the SFG spectra based on the following equation,
combined with Eqs.~(\ref{P2_3}), (\ref{E_IR}), and (\ref{etalon}), and varied time delays $\tau$.
\begin{align}
 I_\mathrm{SFG}(\omega;\tau)
 \propto \left|\int_{-\infty}^\infty dt\, P^{(2)}(t;\tau)e^{i\omega t} \right|^2.  \label{I_SFG2}
\end{align}

\begin{figure}
 \includegraphics[width=7cm]{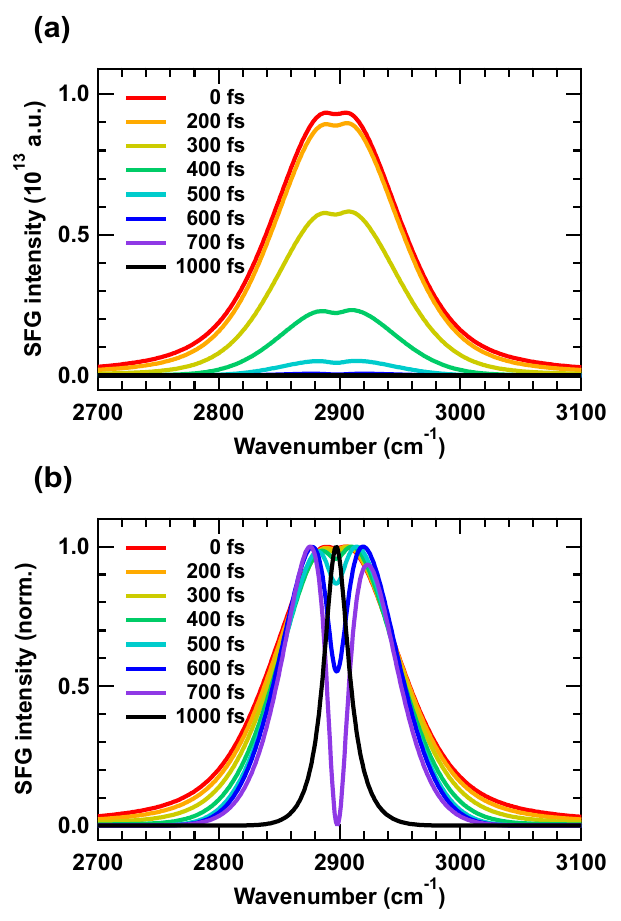}
 \caption{Simulated SFG spectra calculated from Eq.~(\ref{I_SFG2}),
 combined with Eqs.~(\ref{P2_3}), (\ref{E_IR}), and (\ref{etalon}), with the following parameters:
 $A_\mathrm{NR}=2.0\times10^4$, $\theta_\mathrm{NR}=90^\circ$, $A_q=-1$, $\omega_q=2900\,\mathrm{cm}^{-1}$, $\Gamma_q=10\,\mathrm{cm}^{-1}$,
 $\omega_\mathrm{IR}^0=2900\,\mathrm{cm}^{-1}$, $\omega_\mathrm{vis}^0=9680.5\,\mathrm{cm}^{-1}$,
 $\sigma_\mathrm{IR}=\sigma_\mathrm{vis}=40\,\mathrm{cm}^{-1}$, $R=0.93$, and $d=18.6\,\mu\mathrm{m}$.
 $\epsilon_0$ was set to 1.
 $E_\mathrm{IR}^0$ in Eq.~(\ref{E_IR}) and $E_\mathrm{vis}^0$ in Eq.~(\ref{etalon}) were determined
 so that the maximum amplitude of $E_\mathrm{IR}(t)$ and $E_\mathrm{vis}^\mathrm{etalon}(t;\tau)$ was equal to 1, respectively.
 The time delay $\tau$ was varied between 0 and 1000 fs (0, 200, 300, 400, 500, 600, 700, and 1000 fs).
 The spectra shown in panel (b) are normalized to their respective maximum values.
 With increasing $\tau$, the SFG intensity decreases (a),
 while the dip structure arising from the interference between the vibrational response and the NRB becomes more pronounced (b).
 At $\tau=1000\,\mathrm{fs}$, the NRB contribution is almost entirely suppressed, and only the resonant vibrational signal remains.}
 \label{fig3}
\end{figure}

The results show that increasing the time delay reduces the SFG intensity (Fig.~\ref{fig3}(a)),
whereas the dip structure arising from the interference between the vibrational response and the NRB becomes more pronounced (Fig.~\ref{fig3}(b)).
These results demonstrate that an appropriate choice of the time delay enables the NRB to serve as an internal reference.

\begin{figure}
 \includegraphics[width=7cm]{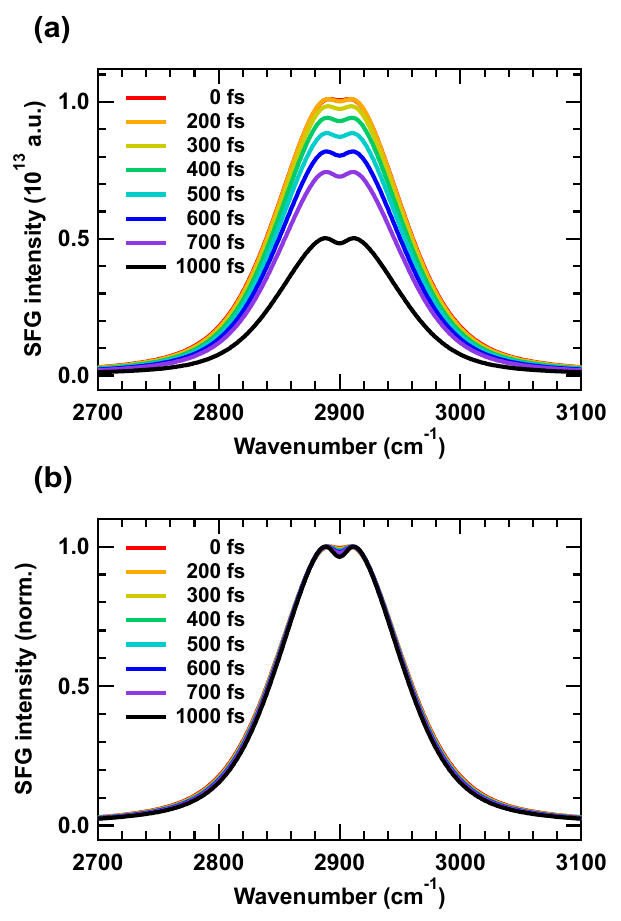}
 \caption{Simulated SFG spectra calculated from Eq.~(\ref{I_SFG2}),
 combined with Eqs.~(\ref{P2_3}), (\ref{E_IR}), and (\ref{E_vis}), with the following parameters:
 $A_\mathrm{NR}=2.0\times10^4$, $\theta_\mathrm{NR}=90^\circ$, $A_q=-1$, $\omega_q=2900\,\mathrm{cm}^{-1}$, $\Gamma_q=10\,\mathrm{cm}^{-1}$,
 $\omega_\mathrm{IR}^0=2900\,\mathrm{cm}^{-1}$, $\omega_\mathrm{vis}^0=9680.5\,\mathrm{cm}^{-1}$,
 $\sigma_\mathrm{IR}=40\,\mathrm{cm}^{-1}$, and $\sigma_\mathrm{vis}=5\,\mathrm{cm}^{-1}$.
 $\epsilon_0$ was set to 1, and
 the time delay $\tau$ was varied between 0 and 1000 fs (0, 200, 300, 400, 500, 600, 700, and 1000 fs).
 The spectra shown in panel (b) are normalized to their respective maximum values.
 With increasing $\tau$, the SFG intensity decreases (a),
 but the relative depth of the dip remains almost the same (b).}
 \label{fig4}
\end{figure}

For comparison, we also calculated the SFG spectra for the case in which the visible pulse had a Gaussian shape
given by Eq.~(\ref{E_vis}), rather than by Eq.~(\ref{etalon}).
The results show that the SFG intensity decreases with increasing time delay (Fig.~\ref{fig4}(a)),
but the relative depth of the dip arising from the interference remains almost the same (Fig.~\ref{fig4}(b)).
This is in sharp contrast to the results shown in Fig.~\ref{fig3}(b),
clearly demonstrating the effectiveness of the temporally asymmetric pulse created by the Fabry–P\'{e}rot etalon.

\subsection{Signal enhancement mechanism in TE-SFG}

In the previous sections, we introduced the general formalism of SFG spectroscopy in the context of far-field observations.
In this subsection, we describe the signal enhancement mechanism in TE-SFG and show that the same formalism also applies here.
As discussed in our previous studies,\cite{Takahashi_JPCL23, Sakurai_NL25} the TE-SFG process consists of three steps:
(i) near-field enhancement of both IR and visible incident light in the nanogap;
(ii) generation of second-order polarization by the enhanced IR and visible near-fields;
(iii) emission of sum-frequency radiation from the induced polarization.
In this subsection, the tilde notation,
typically used to indicate functions in the frequency domain, is omitted for clarity.

In the first step, the incident field is enhanced by plasmonic effects within the nanogap.
This enhancement is quantified by the factor $K(\omega)$,
defined as the ratio of the enhanced near-field $E_\mathrm{NF}(\omega)$ to the incident field $E_0(\omega)$:
\begin{equation}
 E_\mathrm{NF}(\omega) = K(\omega)E_0(\omega).
\end{equation}
This enhanced near-field generates the second-order polarization $P^{(2)}(\omega_\mathrm{SFG})$, given by
\begin{align}
 &P^{(2)}(\omega_\mathrm{SFG})
 = \epsilon_0 \chi^{(2)}(\omega_\mathrm{SFG};\omega_\mathrm{vis},\omega_\mathrm{IR})
 E_\mathrm{NF}(\omega_\mathrm{vis})E_\mathrm{NF}(\omega_\mathrm{IR})  \notag \\
 &= \epsilon_0 \chi^{(2)}(\omega_\mathrm{SFG})
 \left[K(\omega_\mathrm{vis}) E_0(\omega_\mathrm{vis})\right]
 \left[K(\omega_\mathrm{IR}) E_0(\omega_\mathrm{IR})\right].
\end{align}
The induced polarization leads to the emission of a sum-frequency field, expressed as
\begin{equation}
 E_\mathrm{SFG}(\omega_\mathrm{SFG})
  \propto L(\omega_\mathrm{SFG})P^{(2)}(\omega_\mathrm{SFG}),
\end{equation}
where $L(\omega)$ represents the radiation efficiency of the TE-SFG field from $P^{(2)}(\omega)$.
The intensity of the resultant sum-frequency light $I(\omega_\mathrm{SFG})$ in TE-SFG is given by
\begin{align}
 &I(\omega_\mathrm{SFG})
 \propto |E_\mathrm{SFG}(\omega_\mathrm{SFG})|^2  \notag \\
 &\propto |L(\omega_\mathrm{SFG})|^2 |K(\omega_\mathrm{vis})|^2 |K(\omega_\mathrm{IR})|^2
 I_0(\omega_\mathrm{vis}) I_0(\omega_\mathrm{IR}),  \label{I_SFG_enh}
\end{align}
where $I_0(\omega)$ represents the intensity of the incident field ($I_0(\omega)\propto|E_0(\omega)|^2$).
The comparison of Eqs.~(\ref{E_SFG}) and (\ref{I_SFG_enh}) shows that the SFG enhancement is governed
by the $|L(\omega_\mathrm{SFG})|^2|K(\omega_\mathrm{vis})|^2|K(\omega_\mathrm{IR})|^2$ factor,
while the second-order susceptibility $\chi^{(2)}(\omega_\mathrm{SFG})$ is independent of the electric field enhancement.
It should be noted that, if the electric field gradient becomes comparable to the molecular dimensions within the nanogap,
higher-order multipolar contributions may become significant
and should be incorporated into the description of the light–matter interaction.\cite{Morita_book, Fang_PCCP15, Hirano_JCP24}
However, under our experimental conditions, such contributions are negligible, as explained elsewhere.\cite{Takahashi_arXiv1}
Therefore, the TE-SFG spectra can be interpreted in a manner similar to far-field observations,
because the overall process remains the same except for the electric field enhancement.

\section{Experimental Details\label{sec3}}
\subsection{Tip and sample preparation}

\begin{figure}
 \includegraphics[width=\linewidth]{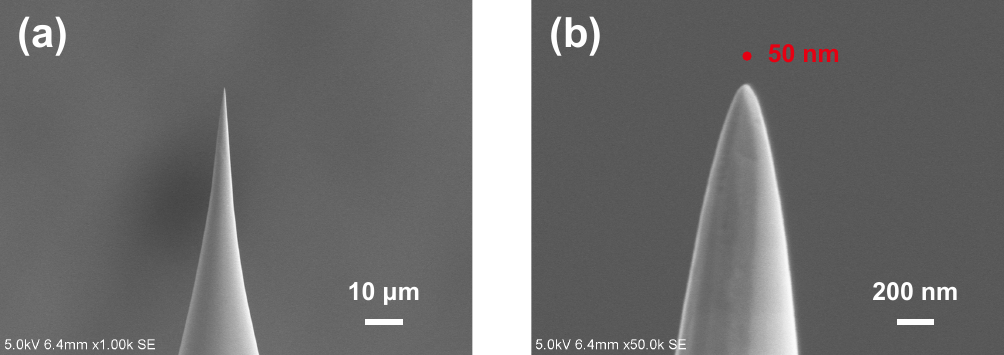}
 \caption{SEM images at different magnifications of STM tip used in this study.
 The red dot in panel (b) represents a 50 nm-diameter circle,
 whose size is comparable to that of the tip apex.}
 \label{fig5}
\end{figure}

\begin{figure}
 \includegraphics[width=6cm]{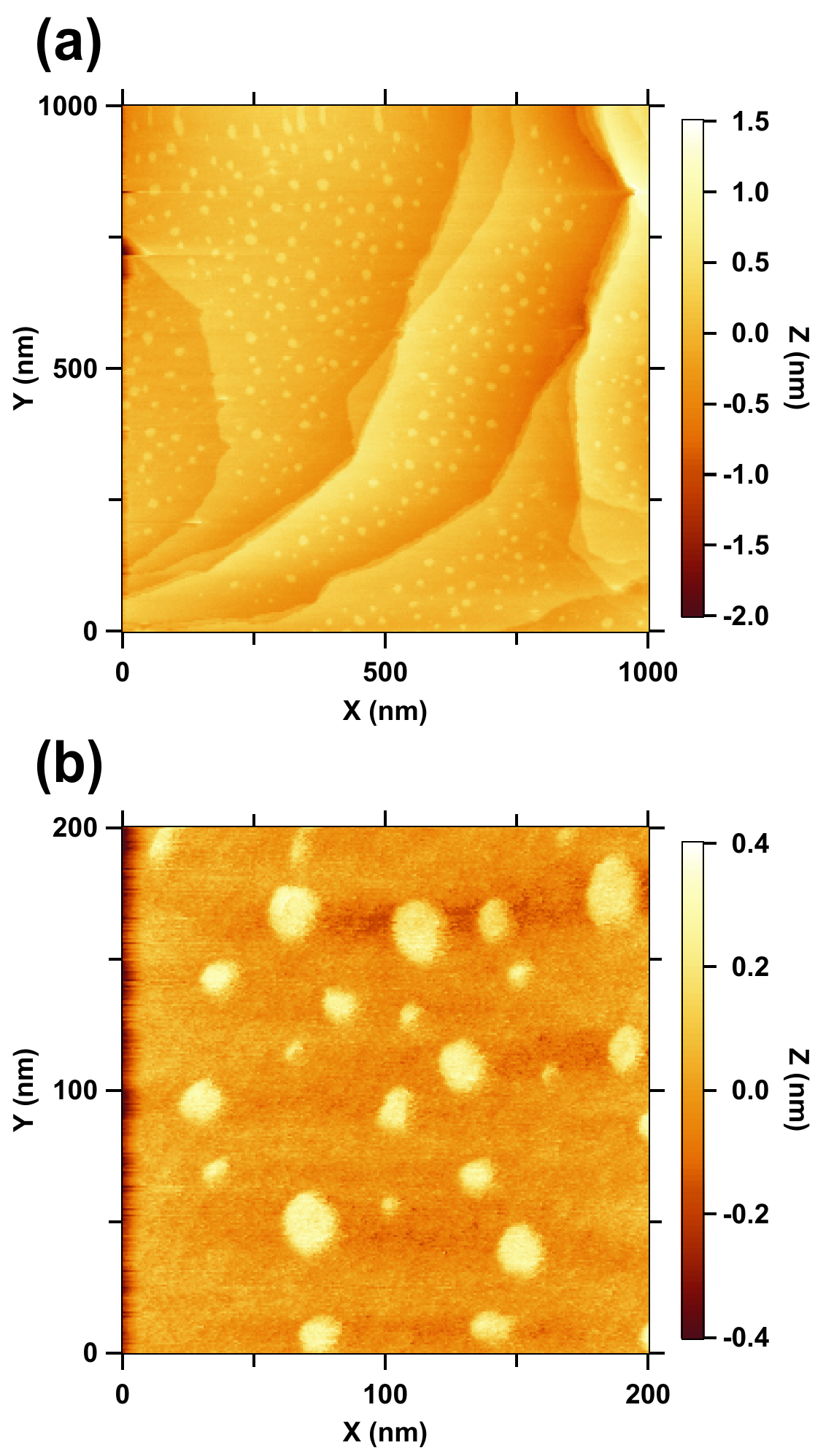}
 \caption{STM images at different magnifications of a 4-MBT SAM on Au(111).
 Color scale bars indicate height.
 The tunneling current was set at 1 nA at a bias voltage of 0.1 V.
 The image sizes are (a) $1\,\mu\mathrm{m}\times1\,\mu\mathrm{m}$ (256 × 256 pixels) and
 (b) $200\,\mathrm{nm}\times200\,\mathrm{nm}$.}
 \label{fig6}
\end{figure}

A homemade STM tip was fabricated by electrochemically etching a gold wire (\O0.25 mm, Nilaco).\cite{Kazuma_JPCC18}
Fig.~\ref{fig5}(a,b) displays scanning electron microscopy (SEM) images of the tip used in this study at differenct magnifications.
The red dot in Fig.~\ref{fig5}(b) represents a 50 nm-diameter circle,
whose size is comparable to that of the tip apex.
The sample was a self-assembled monolayer (SAM) of 4-methylbenzenethiol (4-MBT, Sigma-Aldrich)
on a 300 nm-thick Au film evaporated onto a mica substrate (Unisoku).
The substrate was flame-annealed to obtain an atomically flat Au(111) surface,
and subsequently immersed in a 1 mM ethanol solution of 4-MBT for 24 h,
after which it was rinsed with pure ethanol.
Typical STM images of a 4-MBT SAM are presented in Fig.~\ref{fig6}.
The Au(111) steps in Fig.~\ref{fig6}(a) indicate an atomically flat surface.
In addition, the island-like structures (adatom islands)\cite{Yokota_JPCC20} in Fig.~\ref{fig6}(b),
characteristic of arenethiol SAMs on Au(111),
confirm the formation of a single monolayer of 4-MBT.

\begin{figure*}
 \includegraphics[width=14cm]{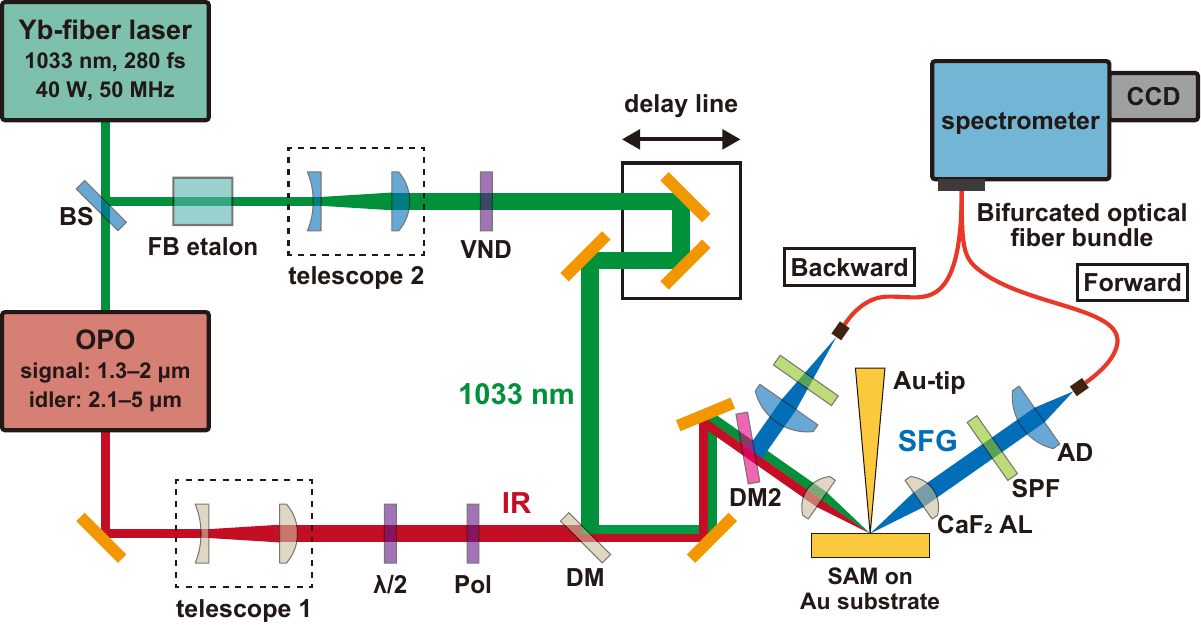}
 \caption{Experimental setup of TE-SFG system.
 BS: beam splitter, OPO: optical parametric oscillator,
 $\lambda$/2: half-wave plate, Pol: polarizer, FB etalon: air-spaced Fabry–P\'{e}rot etalon,
 VND: variable neutral density filter, DM: dichroic mirror, CaF$_2$ AL: CaF$_2$ aspherical lens,
 SPF: short-pass filter, AD: achromatic doublet.
 An additional dichroic mirror (DM2) and a bifurcated optical fiber bundle were employed
 to simultaneously detect both forward- and backward-scattered signals.}
 \label{fig7}
\end{figure*}

\subsection{Experimental setup of TE-SFG system\label{experimental_setup}}

The experimental setup of the TE-SFG system is schematically illustrated in Fig.~\ref{fig7}.
In this system, the output from an amplified Yb-fiber laser (1033 nm, 280 fs, 40 W, 50 MHz; Monaco-1035-40-40, Coherent)
was divided into two arms by a beam splitter.
The first arm served to pump a commercially available optical parametric oscillator (Levante IR, APE),
which generated two output beams: a signal (1.3–2 $\mu$m) and an idler (2.1–5 $\mu$m).
The idler pulses were used to resonantly excite the CH stretching vibrational modes of 4-MBT adsorbed on a gold surface.
The idler beam passed through telescope 1 (two CaF$_2$ lenses with f = $-100$ and 150 mm) to expand the beam diameter
and compensate for its divergence, and subsequently through a half-wave plate (WPLH05M-3500, Thorlabs)
and a polarizer (WP25M-IRA, Thorlabs) to adjust its intensity.
The second arm of the Yb-fiber laser output was directed through an air-spaced Fabry–P\'{e}rot etalon
($R=0.93$, $d=18.6\,\mu\mathrm{m}$; SLS Optics) to narrow the spectral bandwidth and make the pulses temporally asymmetric.
Telescope 2 (two fused silica lenses with f = $-100$ and 150 mm) was employed for the same purpose as telescope 1,
and a variable neutral density filter was used to regulate the beam intensity.

The two laser beams were then combined collinearly at a dichroic mirror
(reflective at 1033 nm and transmissive over the 1300–7000 nm range; custom-made, Sigma Koki)
and simultaneously focused onto the nanojunction formed between the gold tip and sample substrate
in an STM system (USM1400, Unisoku) at an incident angle of 55$^\circ$.
The temporal overlap between the two pulses was controlled by a delay line in the second arm (1033 nm).
Both beams were linearly polarized along the tip axis (\textit{p}-polarization) to effectively excite the gap-mode plasmon.
The emitted sum-frequency signal was collected at a reflection angle of 55°.
Two identical CaF$_2$ aspherical lenses
(effective diameter: 13 mm, f = 19.2, 19.3, and 20.0 mm at $\lambda$ = 800, 1033, and 3500 nm, respectively;
custom-made, Natsume Optical Corporation)
were employed to focus the incident beams and collect the emitted sum-frequency light.
The collected signal was collimated by the CaF$_2$ aspherical lens,
transmitted through a short-pass filter (transmissive over the 500–930 nm range; FESH0950, Thorlabs),
and subsequently focused by an achromatic doublet (f = 50 mm; AC254-050-AB, Thorlabs)
into a multimode optical fiber (\O200 $\mu$m, NA = 0.22; M122L05, Thorlabs).
The fiber-coupled signal was spectrally dispersed by a spectrometer
(Kymera 328i, Andor, with an 830 lines/mm grating blazed at 820 nm or a 300 lines/mm grating blazed at 1000 nm)
and detected using an electronically cooled CCD camera (iDus416, Andor, $2000\times256$ pixels).

Furthermore, an additional dichroic mirror
(reflective over 730–920 nm and transmissive at 1033 nm as well as 2500–8000 nm; custom-made, Lattice Electro Optics)
was placed in the incident beam path (Fig.~\ref{fig7})
to collect the backward-scattered signal.
The backward-scattered sum-frequency signal was subsequently focused by a lens (f = 100 mm; LA1050-AB, Thorlabs),
passed through a short-pass filter (FESH0950),
and coupled into one branch of a bifurcated optical fiber bundle (custom-made, Spectraconn),
after which both the forward- and backward-scattered signals were recorded simultaneously.
The detection results for both forward- and backward-scattered signals are presented in Sec.~\ref{forward_backward}.

The tip–sample distance was regulated by the STM system through feedback control of a constant tunneling current.
Although our STM system can operate under ultrahigh vacuum and cryogenic conditions,
all experiments were performed in ambient air at room temperature.

\subsection{Temporal profile of asymmetric visible pulse\label{temp_prof_vis_pulse}}

\begin{figure}
 \includegraphics[width=8cm]{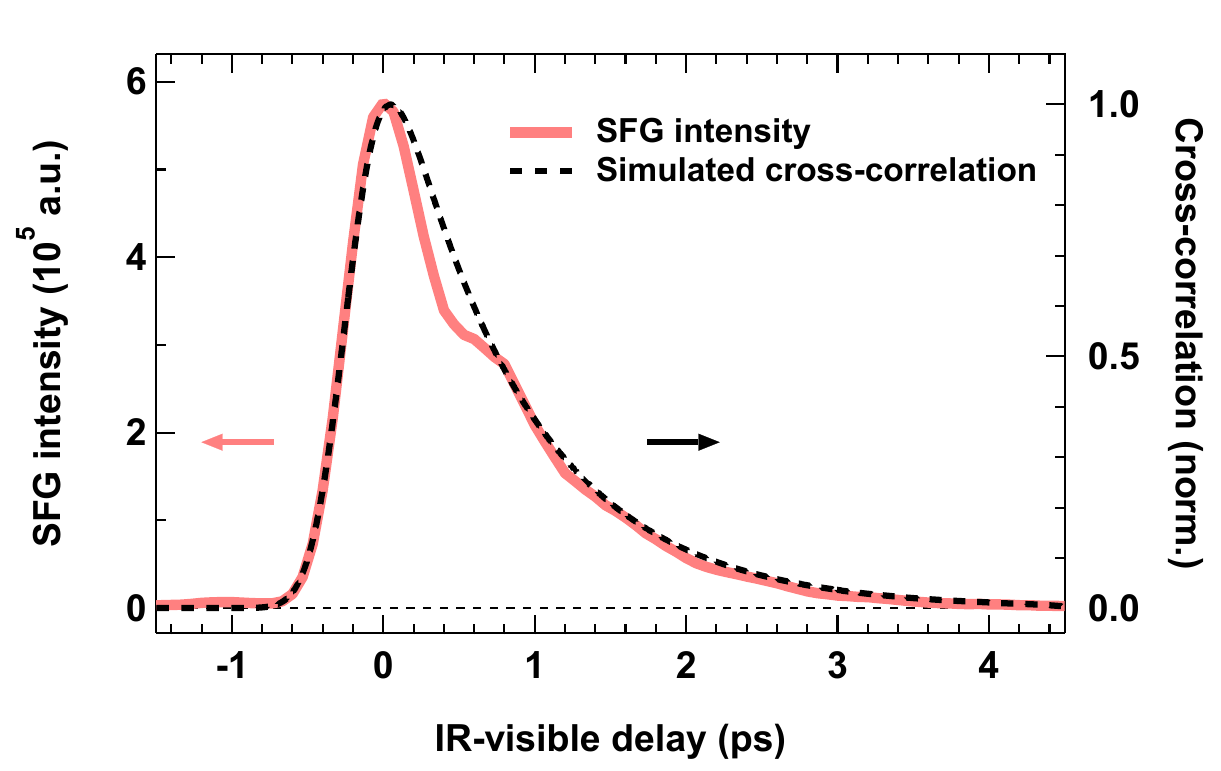}
 \caption{Cross-correlation between temporally asymmetric visible pulse generated by an air-spaced Fabry–P\'{e}rot etalon
 and IR pulse.
 The light red curve shows the experimentally measured cross-correlation obtained from the SFG intensity
 recorded in the far-field geometry on a GaAs substrate (left axis).
 The black dashed curve represents the simulated cross-correlation calculated using Eq.~(\ref{etalon}) (right axis).}
 \label{fig8}
\end{figure}

The temporal profile of the asymmetric visible pulse generated
by an air-spaced Fabry–P\'{e}rot etalon was characterized via cross-correlation with an IR pulse.
In this measurement, the IR and visible pulses were collinearly incident on the GaAs substrate mounted on the STM head,
while recording the SFG intensity at varying time delays.
A representative SFG spectrum is presented in Fig.~\ref{fig14} of Appendix~\ref{app_GaAs}.
During the measurement, the GaAs substrate was retracted by more than 1 $\mu$m from the STM tip to avoid plasmonic enhancement.
The results are shown in Fig.~\ref{fig8} (light red curve),
where the overlaid black dashed curve represents a simulated cross-correlation,
calculated as described below:
\begin{equation}
 I(\tau) = \int_{-\infty}^\infty dt\, \left|E_\mathrm{IR}(t) E_\mathrm{vis}^\mathrm{etalon}(t;-\tau) \right|^2.
\end{equation}
Here, the IR pulse is assumed to be Gaussian, with a FWHM duration of 0.28 ps and a central wavelength of 3500 nm.
$E_\mathrm{vis}^\mathrm{etalon}(t;-\tau)$ was calculated using Eq.~(\ref{etalon}), with an input pulse FWHM of 0.28 ps,
a central wavelength of 1033 nm, a reflectivity $R=0.93$, and an air gap $d=18.6\,\mu\mathrm{m}$.

Overall, the two curves are in good agreement, with two small differences:
a bend in the exponential decay shoulder and a small ($\sim$1\%) intensity bump preceding the steep rise.
These features were also observed in a previous study.\cite{Benderskii_JPCL12}
While their origin is unclear, they appear to be characteristic of a temporally asymmetric pulse generated by a Fabry–P\'{e}rot etalon.
Notably, the simulated cross-correlation shows nearly the same temporal profile
as the calculated envelope of the asymmetric pulse, $|E_\mathrm{vis}^\mathrm{etalon}(t;\tau=0)|^2$
(see Fig.~\ref{fig15} of Appendix~\ref{app_time_envelope});
hence, the observed cross-correlation directly reflects the characteristics of the asymmetric visible pulse.

\section{Results and Discussion\label{sec4}}
\subsection{Acquisition of TE-SFG spectra}

A tunable IR pulse (central wavenumber: 2850–3000 cm$^{-1}$)
and a temporally asymmetric narrowband visible pulse (1033 nm, with a FWHM of 10 cm$^{-1}$)
were focused onto the nanojunction formed between the STM tip and the sample substrate.
The input powers of the IR and 1033 nm beams were 20 and 0.2 mW, respectively, at a repetition frequency of 50 MHz.
The average (peak) power densities at the focus were estimated to be
$2.9 \times 10^3$ ($1.9 \times 10^8$) and $3.6 \times 10^2$ ($7.1 \times 10^6$) W/cm$^2$ for the IR and 1033 nm pulses,
assuming pulse durations of 300 fs and 1 ps, respectively.
The IR–visible pulse delay was set to the value
where the SFG intensity from GaAs reached its maximum;
this delay was defined as $\tau=0$ (Fig.~\ref{fig8}).

\begin{figure}
 \includegraphics[width=8cm]{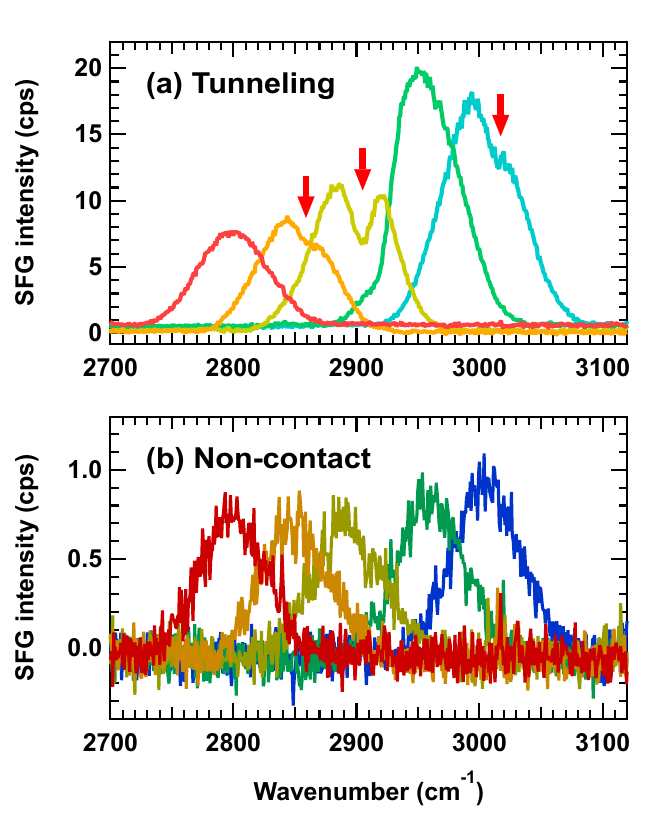}
 \caption{Comparison of SFG spectral intensities under two different conditions:
 (a) tunneling regime, with the substrate positioned very close to the tip ($\lesssim$1 nm) (2 nA, 0.05 V),
 and (b) non-contact regime, with the substrate separated from the tip by 50 nm.
 Strong signals originating from tip enhancement appear in (a),
 whereas only weak far-field signals are observed in (b).
 Exposure time: 10 min per spectrum; intensities normalized to counts per second (cps).
 The three red arrows mark spectral dips resulting from vibrational resonance.
 The ratio of the tip-enhanced to far-field signals was in the range of 13–27,
 based on the integrated intensities in (a) and (b).}
 \label{fig9}
\end{figure}

By tuning the IR wavenumber,
we acquired five SFG spectra for each of the two experimental conditions: tunneling and non-contact.
In the tunneling regime,
the substrate was positioned very close to the tip ($\lesssim$1 nm),
with the current set to 2 nA at a bias voltage of 0.05 V,
and strong SFG signals were observed (Fig.~\ref{fig9}(a)).
In contrast, only very weak signals, attributed to far-field effects, were detected
in the non-contact regime, with the substrate retracted by 50 nm from the tip (Fig.~\ref{fig9}(b)).
This strong contrast in SFG intensities
indicates that the SFG spectra observed in the tunneling regime originate from tip enhancement.
Such enhancement arises from two synergistic effects: the antenna effect,
which focuses IR light onto the tip apex, and plasmonic enhancement within the nanogap,
which enhances the SFG emission efficiency in the visible region,
as reported in previous studies.\cite{Takahashi_JPCL23, Sakurai_NL25}
In the tunneling regime, spectral dips were observed at the positions marked by the three red arrows (Fig.~\ref{fig9}(a)).
These features result from interference between vibrationally resonant and non-resonant contributions.
In the non-contact regime, a dip appears to be present in the dark yellow line,
but its structure is not clear, owing to the low signal-to-noise ratio.
Finally, the ratio of tip-enhanced to far-field signals was calculated to be in the range of 13–27
from the integrated SFG intensities obtained in the tunneling and non-contact regimes.
Based on these values, the signal enhancement factor of TE-SFG is discussed in Sec.~\ref{signal_enhancement}.

In our STM configuration, the sample is scanned instead of the tip,
which ensures that the incident laser beams remain focused on the tip apex throughout the scan.
During SFG measurements in the tunneling regime, the sample was scanned over a $3\,\mu\mathrm{m}\times3\,\mu\mathrm{m}$ area
at a speed of 2.5 $\mu$m/s to avoid sample damage.
Measurements were repeated in the same region,
and the STM images obtained with and without laser irradiation of the same scanned area
are compared in Fig.~\ref{fig16} of Appendix~\ref{STM_images}.

The focal spot diameters for the 1033 and 3500 nm beams were estimated to be 8.5 and 29.7 $\mu$m, respectively,
using the relation $2r = 4\lambda f/\pi D$, where $r$ is the spot radius, $\lambda$ is the wavelength,
$f$ is the focal length of the lens, and $D$ is the incident beam diameter (3 mm).\cite{Saleh_book}
The depth of focus is given by $2z_\mathrm{R}=2\pi r^2/\lambda$, where $z_\mathrm{R}$ is the Rayleigh length;\cite{Saleh_book}
using this relationship, the depths of focus for 1033 and 3500 nm light were calculated to be 108 and 396 $\mu$m, respectively.
Considering these values, retracting the substrate by only 50 nm from the tip
or scanning the sample during the measurement
would not have a significant effect,
because the focal spot could still be regarded as remaining in the same position.

\subsection{Effect of IR–visible pulse delay on TE-SFG response}

\begin{figure*}
 \includegraphics[width=12cm]{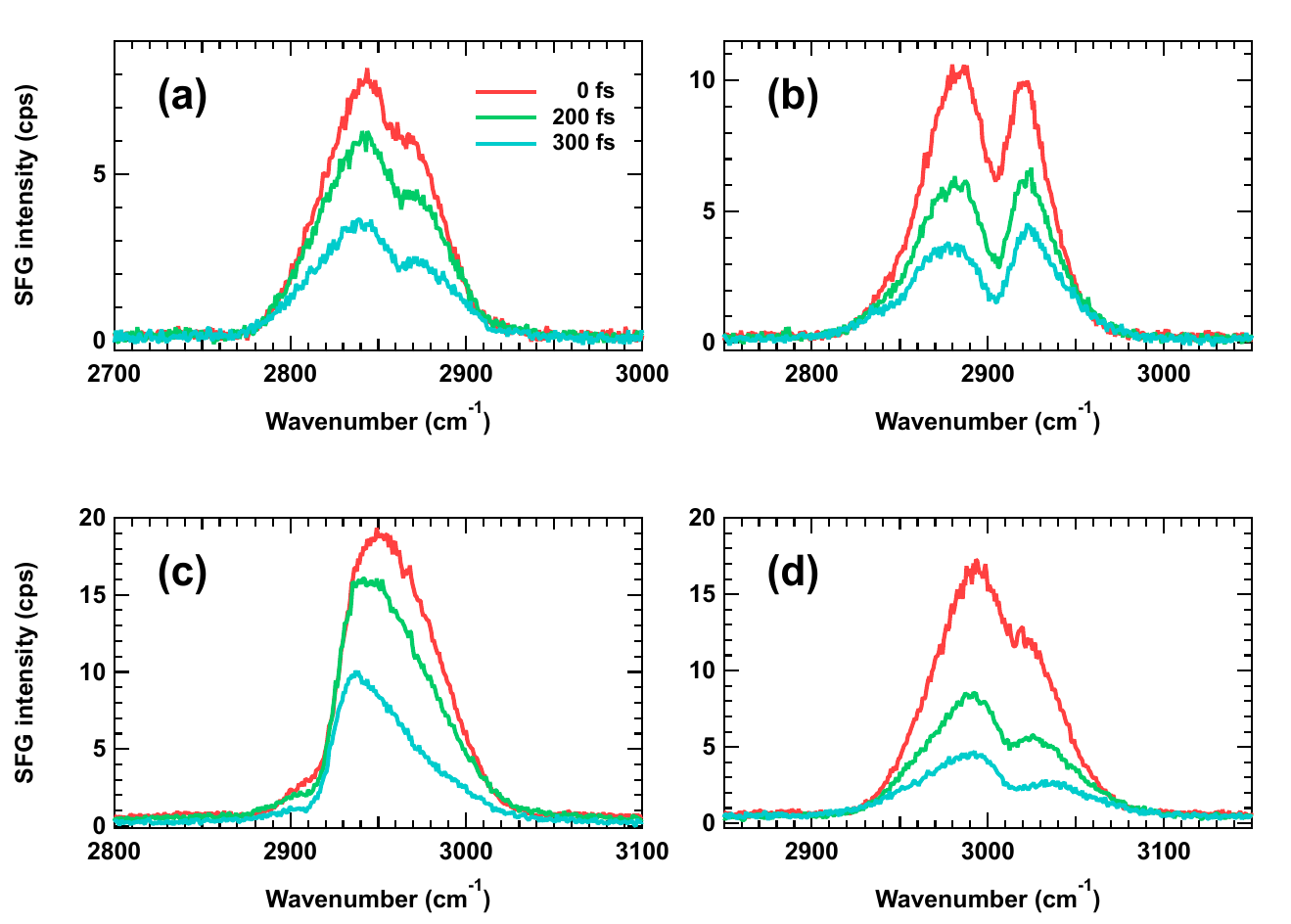}
 \caption{Tip-enhanced SFG spectra measured at four IR wavenumbers
 for IR–visible pulse delays of 0, 200, and 300 fs.
 To remove background contributions, spectra acquired in the non-contact regime were subtracted
 from those recorded in the tunneling regime.}
 \label{fig10}
\end{figure*}

We investigated the effect of the IR–visible pulse delay on the TE-SFG response.
The delay $\tau$ was set to 200 and 300 fs, as well as 0 fs, relative to the spectra shown in Fig.~\ref{fig9}.
Fig.~\ref{fig10} shows the corresponding SFG spectra obtained for four central IR wavenumbers.
To remove background contributions, primarily arising from far-field effects,
the spectra obtained in the non-contact regime were subtracted from those recorded in the tunneling regime.
As the delay increased, the overall spectral intensity decreased (Fig.~\ref{fig10}(a–d)),
whereas the dip structures became more pronounced,
as most clearly observed in Fig.~\ref{fig10}(d).
This trend closely resembles that of the simulated SFG spectra shown in Fig.~\ref{fig3}.
These results demonstrate that interferometric detection
using a temporally asymmetric pulse is an effective approach for TE-SFG spectroscopy.

\begin{figure*}
 \includegraphics[width=16cm]{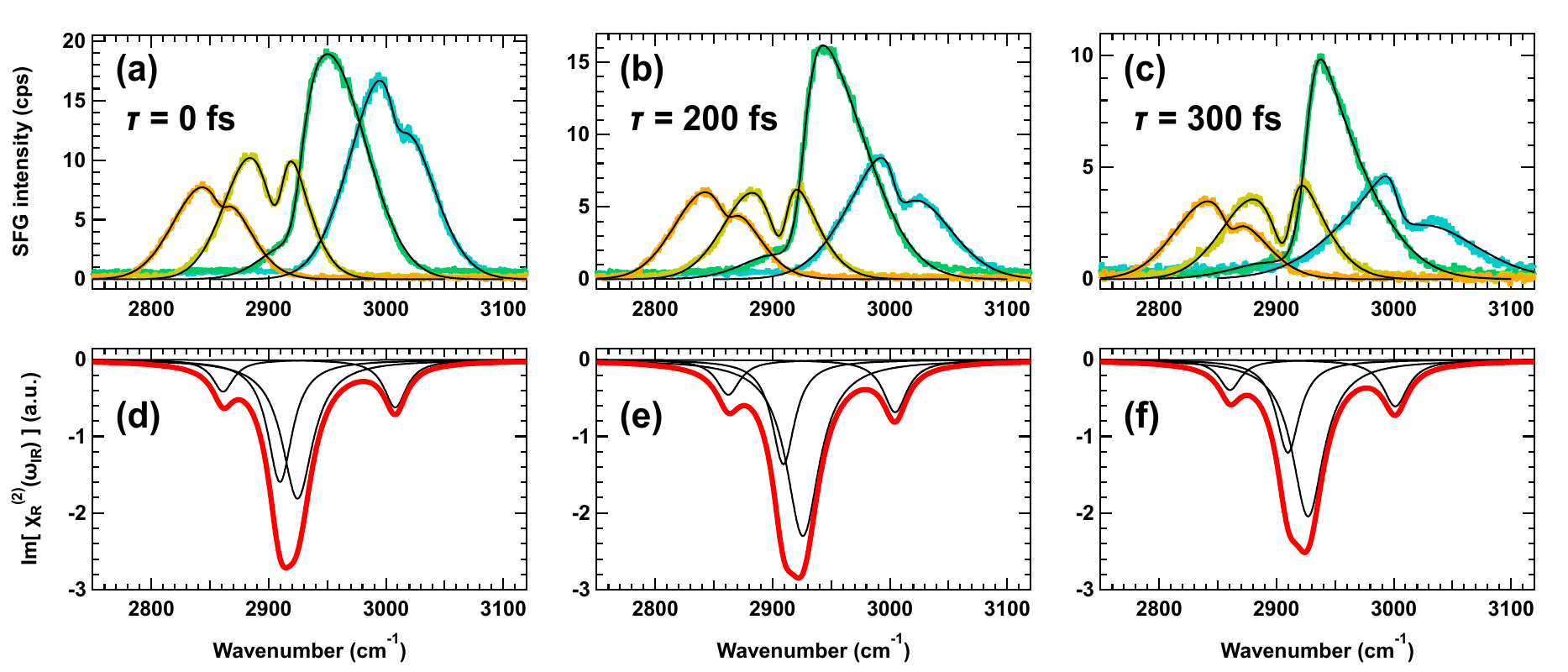}
 \caption{TE-SFG spectra for IR–visible pulse delays of (a) 0, (b) 200, and (c) 300 fs.
 The corresponding fitting curves based on Eq.~(\ref{I_SFG3}) are shown as black lines.
 The respective imaginary parts of the second-order susceptibility ($\mathrm{Im}\left[\chi^{(2)}_\mathrm{R}(\omega_\mathrm{IR})\right]$)
 for IR–visible pulse delays of (d) 0, (e) 200, and (f) 300 fs are shown as black lines.
 The red curves represent the sum of these contributions.}
 \label{fig11}
\end{figure*}

To extract information on molecular vibrations,
the TE-SFG spectra were fitted using the following equation, based on Eq.~(\ref{I_SFG}),
assuming a single vibrational mode within each spectral range:
\begin{align}
 I_\mathrm{SFG}(\omega)
 =& \left|A_\mathrm{NR}e^{i\theta_\mathrm{NR}} + \frac{A_q}{\omega_q - \omega_\mathrm{IR} - i\Gamma_q} \right|^2  \notag \\
 &\times \exp\left[-\frac{(\omega_\mathrm{IR}-\omega_\mathrm{IR}^0)^2}{\sigma_\mathrm{IR}^2} \right].  \label{I_SFG3}
\end{align}
The time delay effects were not explicitly included in this equation, as they are not needed for spectral assignment.
The phase of the non-resonant term ($\theta_\mathrm{NR}$) was assumed to be $\sim$90$^\circ$,
as reported in a previous study.\cite{Castner_Langmuir10, Hore_JPCC15}
The fitting curves, shown as black lines in Fig.~\ref{fig11}(a–c),
are in good agreement with the experimental data.
Next, the imaginary parts of the vibrationally resonant contribution to the second-order susceptibility
($\mathrm{Im}\left[\chi^{(2)}_\mathrm{R}(\omega_\mathrm{IR})\right]$) were calculated using the same parameters as in the fitting procedure,
as described below:
\begin{align}
 \mathrm{Im}\left[\chi^{(2)}_\mathrm{R}(\omega_\mathrm{IR})\right]
 &= \mathrm{Im}\left[ \frac{A_q}{\omega_q-\omega_\mathrm{IR}-i\Gamma_q} \right]  \notag \\
 &= \frac{A_q\Gamma_q}{(\omega_\mathrm{IR}-\omega_q)^2 + \Gamma_q^2},  \label{Im_chi2}
\end{align}
and plotted as black lines in Fig.~\ref{fig11}(d–f).
These curves represent the purely absorptive response of the $q$-th vibrational mode.
The sum of these contributions is plotted as red lines in Fig.~\ref{fig11}(d–f).

The extracted vibrational mode wavenumbers for $\tau=0\,\mathrm{fs}$ were 2861, 2910, 2925, and 3008 cm$^{-1}$,
and the other fitting parameters are listed in Table~\ref{tab1} of Appendix~\ref{app_fitting_parameter}.
In our previous study,\cite{Sakurai_NL25} three vibrational modes were identified for the same sample (a 4-MBT SAM on Au(111))
using TE-SFG spectroscopy: the methyl symmetric stretching vibration (CH$_3$-SS) at 2853 cm$^{-1}$,
a Fermi resonance (CH$_3$-FR) between the CH$_3$-SS mode and an overtone of the methyl bending vibration at 2904 cm$^{-1}$,
and the degenerate antisymmetric stretching vibration of the methyl group (CH$_3$-DS) at 2939 cm$^{-1}$.
In the present study, the CH$_3$-DS mode appeared at a 14 cm$^{-1}$ lower wavenumber than previously reported,
whereas the CH$_3$-SS and CH$_3$-FR modes were observed at 8 and 6 cm$^{-1}$ higher wavenumbers, respectively.
The opposite shift of the CH$_3$-DS mode is likely due to the dip signal being located off-center relative to the pulse spectrum,
thereby amplifying the fitting error.
Alternatively, the dip may originate from the CH$_3$-FR rather than CH$_3$-DS vibration,
in which case the CH$_3$-DS mode would not have been detected.

\begin{figure*}
 \includegraphics[width=12cm]{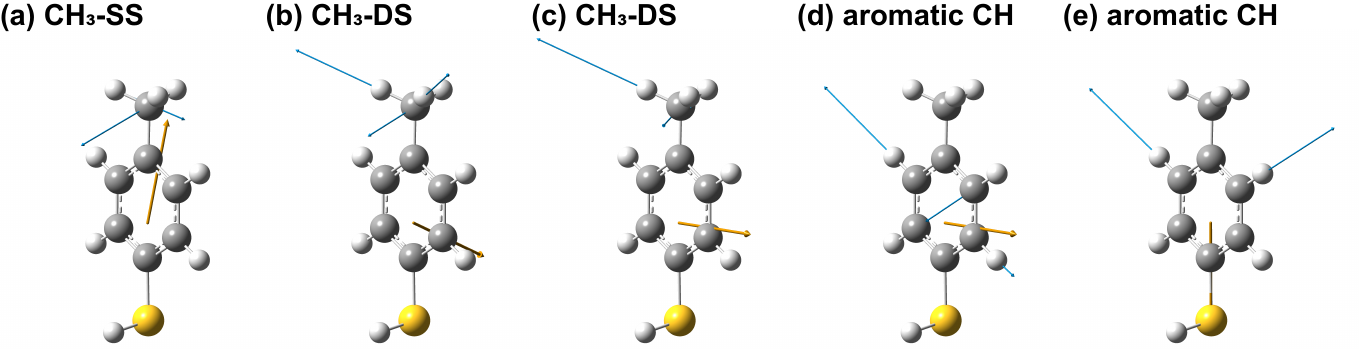}
 \caption{Vibrational modes for 4-MBT displayed using the GaussView software,\cite{GaussView}
 calculated at the B3LYP/6-31g(d) level of DFT using the Gaussian 16 program.\cite{Gaussian}
 The blue and orange vectors indicate the displacements of each atom
 and the dipole moments for each vibrational mode, respectively.
 The CH$_3$-DS mode consists of two degenerate modes, (b) and (c);
 similarly, the aromatic CH mode also comprises two modes, (d) and (e).}
 \label{fig12}
\end{figure*}

Quantum chemical calculations indicate that the dipole moment of the CH$_3$-DS mode is almost perpendicular
to the molecular axis of 4-MBT, as shown in Figs.~\ref{fig12}(b) and \ref{fig12}(c).
Hence, if the 4-MBT molecule is oriented normal to the substrate,
the tip enhancement becomes ineffective,
because the near-field enhancement predominantly arises
along the substrate normal direction.\cite{Takahashi_JPCL23}
The SAM of 4-MBT is known to form multiple phases corresponding to different alignment structures.\cite{Borguet_JPCC07}
This implies that the SAM used in the present study may have a structural order different from that observed previously,
which could account for the disappearance of the CH$_3$-DS mode signal or the shifts in the vibrational wavenumbers.

Next, we examine the fourth vibrational mode at 3008 cm$^{-1}$.
The dip from this mode was very small at $\tau=0\,\mathrm{fs}$, but became pronounced at $\tau=200$ and 300 fs.
This mode corresponds to the CH vibration associated with the aromatic ring
(aromatic CH, as shown in Figs.~\ref{fig12}(d) and \ref{fig12}(e)),
which is known to be weaker than the other three methyl modes;\cite{Dreesen_TSF06}
in fact, it was absent in our previous study.\cite{Sakurai_NL25}
Its detection in this work, and its delay-induced enhancement,
highlight the benefits of using a temporally asymmetric visible pulse,
with an appropriate time delay relative to the IR pulse.
Although the utility of this method has been proven for far-field SFG,
the present results demonstrate its applicability to TE-SFG measurements.
Because TE-SFG probes molecular vibrations within the STM nanojunction, the intrinsic signal is inherently weak.
Furthermore, TE-SFG relies on gap-mode plasmonic enhancement, which unavoidably produces a non-resonant SFG background,
especially when a plasmonic tip and substrate are used.
Introducing a time delay between the two laser pulses provides a way
to balance the weak vibrational response against the strong NRB.

\subsection{Sign of $\bm{\mathrm{Im}\left[\chi^{(2)}_\mathrm{R}(\omega_\mathrm{IR})\right]}$ term and molecular orientation}

Although the assignment of methyl vibrational modes in the present study is slightly different
from that adopted in the previous work,
both consistently exhibit a negative sign of
the $\mathrm{Im}\left[\chi^{(2)}_\mathrm{R}(\omega_\mathrm{IR})\right]$ term,
resulting from the negative sign of $A_q$ in Eq.~(\ref{Im_chi2}).
This term corresponds to the amplitude of the molecular hyperpolarizability, and its tensor form is expressed as
\begin{equation}
 A_{q,lmn} = \frac{1}{2\epsilon_0\omega_q}
  \left(\frac{\partial\alpha_{lm}^{(1)}}{\partial Q_q}\right)
  \left(\frac{\partial\mu_{n}}{\partial Q_q}\right),
\end{equation}
where $\{l,m,n\}$ are indices corresponding to the molecular-frame Cartesian coordinates $\{a,b,c\}$,
while $\alpha^{(1)}$ and $\mu$ denote the Raman polarizability and IR transition dipole moment, respectively,
and $Q_q$ is the normal coordinate of the $q$-th vibrational mode.\cite{Shen_CPL90, Wang_JPCB04}

Following the definition of the molecular frame in the previous study by Jena \textit{et al.},\cite{Hore_JPCC11}
the local $c$-axis of the methyl group is defined as pointing from the methyl carbon toward the methyl hydrogens.
For the CH$_3$-SS mode at the end of an alkyl chain,
the polarizability derivative is $\partial\alpha_{cc}^{(1)}/\partial Q_q<0$,
while the dipole moment derivative is $\partial\mu_c/\partial Q_q>0$,\cite{Hore_JPCC11}
which results in $\mathrm{Im}\left[\chi^{(2)}_\mathrm{R}(\omega_\mathrm{IR})\right]<0$.
In addition, for the CH$_3$-SS mode, the tensor component $A_{q,ccc}$ has a significantly larger absolute value
than other hyperpolarizability components.\cite{Takahashi_arXiv1}
This means that only the $A_{q,ccc}$ component needs to be considered,
and the averaging over the hyperpolarizability components
based on the Euler rotational transformation matrix\cite{Shen_PRB99, Morita_book, Shen_CPL90} can be omitted,
provided that the $c$-axis is nearly aligned with the surface normal.
Because the near-field enhancement predominantly arises along the surface normal within the nanogap,
this field selectively excites the vibrational modes oriented along the surface normal.
This is consistent with the adsorption structure of 4-MBT,
in which the terminal methyl groups are oriented with their hydrogen atoms pointing away from the substrate.
This demonstrates that phase-sensitive TE-SFG can determine the absolute molecular orientations
at the nanoscale with high precision.
The correlation between the sign of $\mathrm{Im}\left[\chi_\mathrm{R}^{(2)}(\omega_\mathrm{IR})\right]$ and the molecular orientations for other vibrational modes is discussed elsewhere.\cite{Takahashi_arXiv1}

\subsection{Simultaneous detection of forward- and backward-scattered signals\label{forward_backward}}

\begin{figure}
 \includegraphics[width=8cm]{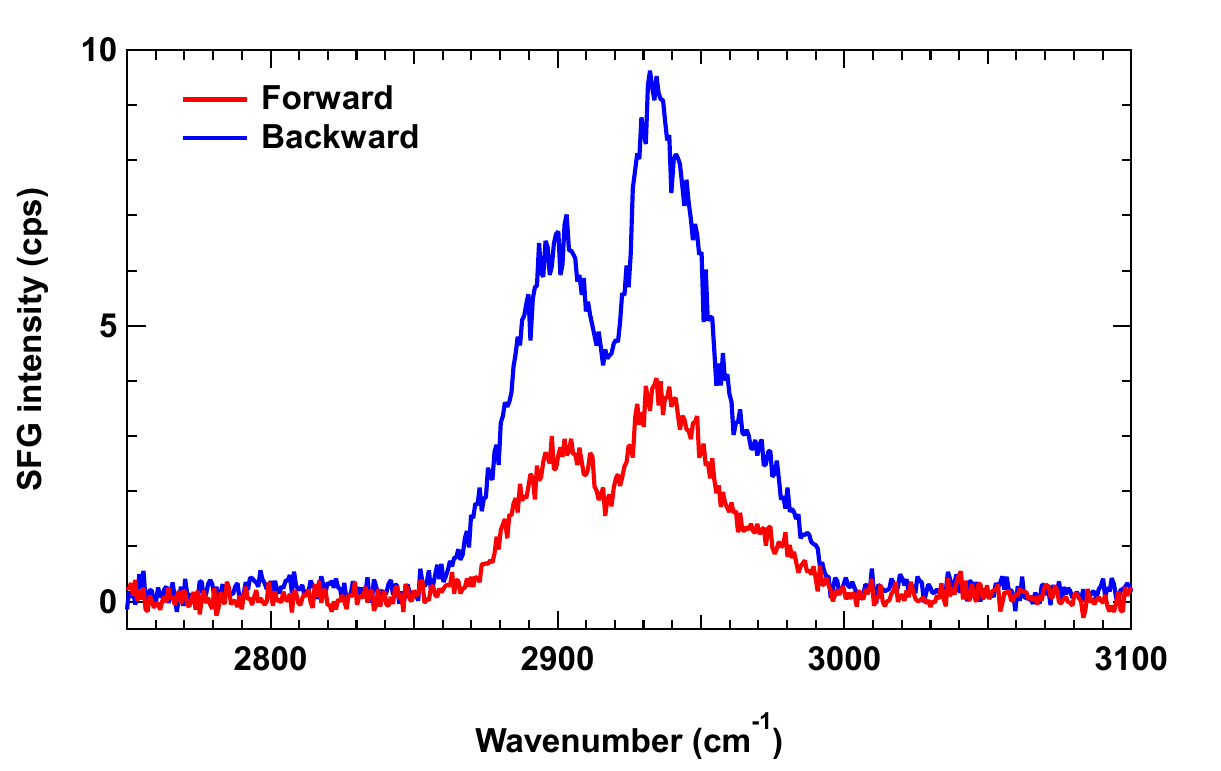}
 \caption{Forward-scattered (red) and backward-scattered (blue) signals
 simultaneously detected in TE-SFG spectroscopy measurements.
 The observed dip corresponded to the CH$_3$-SS mode.
 The input powers of the IR and 1033 nm beams were 6 and 0.2 mW, respectively.
 The tunneling current was set to 245 pA, with a bias voltage of 0.5 V.}
 \label{fig13}
\end{figure}

As discussed in Sec.~\ref{experimental_setup},
we performed the simultaneous detection of forward- and backward-scattered SFG signals;
a representative result is presented in Fig.~\ref{fig13}.
The SFG signals were observed in both forward and backward directions,
and exhibited a dip corresponding to the CH$_3$-DS mode.
This result is in sharp contrast with the coherent SFG signals acquired in a far-field geometry,
which appeared only in the forward direction satisfying the phase-matching condition
(see Fig.~\ref{fig14} of Appendix~\ref{app_GaAs}).
As shown in Fig.~\ref{fig13}, the backward-scattered signal was approximately twice as strong as the forward-scattered signal.
However, the intensity distributions varied significantly with the specific tips used.
This variation is attributed to the nanoscale morphology of the tip apex.
The tip morphology significantly influences the generation of the near-field in the plasmonic nanocavity
and thereby affects both the signal generation efficiency and emission direction,
as previously reported for STM luminescence.\cite{Ham_ApplNanoMater20}
These findings clearly demonstrate that the observed SFG signals originate
from tip enhancement, rather than from a coherent SFG process.

\subsection{Signal enhancement factor of TE-SFG and comparison with surface-enhanced SFG\label{signal_enhancement}}

Because the focal spot of the visible beam (8.5 $\mu$m in diameter) is smaller than that of the IR beam,
the far-field SFG signal can be considered to originate from the former spot.
On the other hand, the TE-SFG signal arises from an area comparable in size to the tip apex
($\sim$50 nm).\cite{Hillenbrand_NL06}
In addition, the coherent SFG signal is emitted only in the forward direction satisfying the phase-matching condition,
whereas the TE-SFG signal is emitted more isotropically, depending on the nanoscale morphology of the tip apex,
as shown in Fig.~\ref{fig13}.
The numerical aperture of the collection lens is 0.34 at 800 nm;
thus, assuming isotropic TE-SFG emission from a point source,
the lens captures only 6.0\% of the radiation within the upper hemisphere.
Finally, the ratio of the tip-enhanced to far-field signals was in the range of 13–27,
as shown in Fig.~\ref{fig9}.
From these values, the signal enhancement factor was estimated to be $6.3\times 10^6 - 1.3\times 10^7$.
This factor is two orders of magnitude greater than that previously reported
by Baldelli \textit{et al.} for surface-enhanced SFG (SE-SFG),
where such enhancement was realized through highly ordered plasmonic structures.\cite{Baldelli_JCP00}

In the case of SE-SFG, when Au or Ag nanoparticles are randomly distributed on the substrate surface,
the resulting enhancement factor rarely exceeds one order of magnitude.\cite{Okuno_CPL15, Linke_JPCC19, Busson_JPCC19}
This difference reflects the coherent nature of SFG:
as fields emitted from hot spots formed at particle–substrate junctions interfere,
ordered arrays promote constructive interference,
whereas random distributions of hot spots average out the phases of the emitted fields and suppress the macroscopic signal.
In comparison, Raman-based techniques,
such as tip-enhanced (TERS) and surface-enhanced (SERS) Raman scattering,
are intrinsically incoherent,
and their signals are simply additive in intensity,
without requiring phase alignment.
Further details are provided in Appendix~\ref{app_SFG_vs_Raman}.
In TE-SFG, the signal enhancement occurs within a single nanogap,
which prevents destructive interference and thus enables strong signal enhancement.

\section{Conclusion\label{sec5}}

We performed TE-SFG spectroscopy measurements on a SAM of 4-MBT adsorbed on a Au surface.
The observed signals exhibited dips arising from vibrational modes within the broadband NRB spectra.
We theoretically formulated and implemented a technical scheme to suppress the NRB
by making the second visible pulse temporally asymmetric and introducing a controlled delay
between the first IR and second visible pulses.
As a result, the contrast between the resonant and non-resonant signals was modulated,
thereby amplifying the vibrational response with increasing delay.
The observed vibrational modes were assigned to the CH stretching of the methyl group,
and we also detected the aromatic CH mode,
which had not been observed in the previous study.\cite{Sakurai_NL25}
Moreover, owing to the interference between vibrational responses and NRB,
TE-SFG became phase-sensitive, demonstrating its ability to determine absolute molecular orientations at the nanoscale.
Furthermore, forward- and backward-scattered signals were simultaneously detected,
confirming that the observed signals originated from tip enhancement rather than far-field contributions.
Finally, the signal enhancement factor was quantitatively estimated to be $6.3\times 10^6 - 1.3\times 10^7$
based on the experimental data.

SFG imaging data were not directly presented in this study;
however, we continued to investigate this subject
and recently achieved spatial resolution on the nanometer scale ($\sim$30 nm).\cite{Takahashi_arXiv1}
Notably, Roelli \textit{et al.} demonstrated TE-SFG using a CW rather than pulsed laser,
which generated no NRB, and observed both SFG and difference-frequency generation (DFG) signals.\cite{Roelli_LSA25}
TE-SFG imaging based on a CW laser may be a promising approach.
In addition, we investigated the effects of a bias voltage applied in the STM nanojunction
on the intensity of the TE-SFG signals.
Our results revealed a significant enhancement of the signal as the applied bias increased.\cite{Takahashi_arXiv2}

Finally, we consider the applicability of TE-SFG to time-resolved spectroscopy.
This method can be used to measure the vibrational dynamics by varying the IR–visible delay.
Recent developments in pulse-excited TERS have advanced this field,\cite{VanDuyne_JPCL14, Kern_NL22}
and ultrafast coherent phonon dynamics has been observed in graphene nanoribbons.\cite{Kern_NatComm23}
Here, the decay of the vibrational response obtained
by varying the IR–visible delay corresponds to the total dephasing time ($T_2$);
however, by extending the setup to a pump–probe scheme (i.e., an IR pump and SFG probe configuration),\cite{Guyot-Sionnest_PRL90}
the population decay time ($T_1$) can also be extracted, where $T_2,T_1$,
and the pure dephasing time ($T_2^*$) are related by $1/T_2=1/2T_1 + 1/T_2^*$.\cite{Ueba_PSS97}
In principle, $T_1$ can be obtained using a degenerate IR pump–probe measurement;
however, this requires the detection of a very small change in the IR probe absorption.
In contrast, SFG is inherently background-free,
because the excitation and detection wavelengths are different.
Therefore, TE-SFG is well suited for detecting weak signals and shows great potential for pump–probe applications.

\begin{acknowledgments}
 We thank the members of the Equipment Development Center of the Institute for Molecular Science (IMS),
 Masaki Aoyama, Takuhiko Kondo, Nobuo Mizutani, and Takuro Kikuchi, for their technical assistance with machine work
 as well as Tomonori Toyoda and Kazunori Kimura for their technical assistance with electronics.
 We also thank Aya Toyama and Osamu Ishiyama at the Instrument Center of IMS for their tutorial on SEM.
 SEM observation was conducted at IMS, supported by
 ``Advanced Research Infrastructure for Materials and Nanotechnology in Japan (ARIM)''
 of the Ministry of Education, Culture, Sports, Science and Technology (MEXT),
 Proposal Number JPMXP1224MS5015.

 T.S. acknowledges financial support from JST-PRESTO (JPMJPR1907), JST-CREST (JPMJCR22L2),
 JSPS KAKENHI Grants-in-Aid for Scientific Research (A) (19H00865 and 22H00296),
 and Grant-in-Aid for Transformative Research Areas (A) (24H02205).
 A.M. acknowledges financial support from JSPS KAKENHI Grants-in-Aid for Scientific Research (A) (20H00368) and (B) (25K01721).
 A.S. acknowledges financial support from JSPS KAKENHI Grants-in-Aid for Scientific Research (B) (23H01855),
 Grants-in-Aid for Challenging Research (Exploratory) (24K21759).
 S.T. acknowledges financial support from Grants-in-Aid for JSPS Fellows (22KJ3099).
\end{acknowledgments}

\section*{AUTHOR DECLARATIONS}

\subsection*{Conflict of Interest}

\noindent
The authors have no conflicts to disclose.

\subsection*{Author Contributions}

\noindent
\textbf{Atsunori Sakurai:} Conceptualization (equal); Methodology (equal); Formal Analysis (equal); Investigation (lead); Writing – Original Draft (lead); Writing – Review \& Editing (equal); Funding Acquisition (supporting).
\textbf{Shota Takahashi:} Methodology (equal); Investigation (supporting); Resources (equal); Writing - Writing – Review \& Editing (equal); Funding Acquisition (supporting).
\textbf{Tatsuto Mochizuki:} Methodology (equal); Investigation (supporting); Resources (equal).
\textbf{Tomonori Hirano:} Formal Analysis (equal); Writing – Review \& Editing (equal).
\textbf{Akihiro Morita:} Formal Analysis (equal); Writing – Review \& Editing (equal); Funding Acquisition (supporting).
\textbf{Toshiki Sugimoto:} Conceptualization (equal); Methodology (equal); Writing – Review \& Editing (equal); Project Adminstration (lead); Funding Acquisition (lead).

\appendix

\section{SFG and 2PFL detection from GaAs\label{app_GaAs}}

\begin{figure}
 \includegraphics[width=8cm]{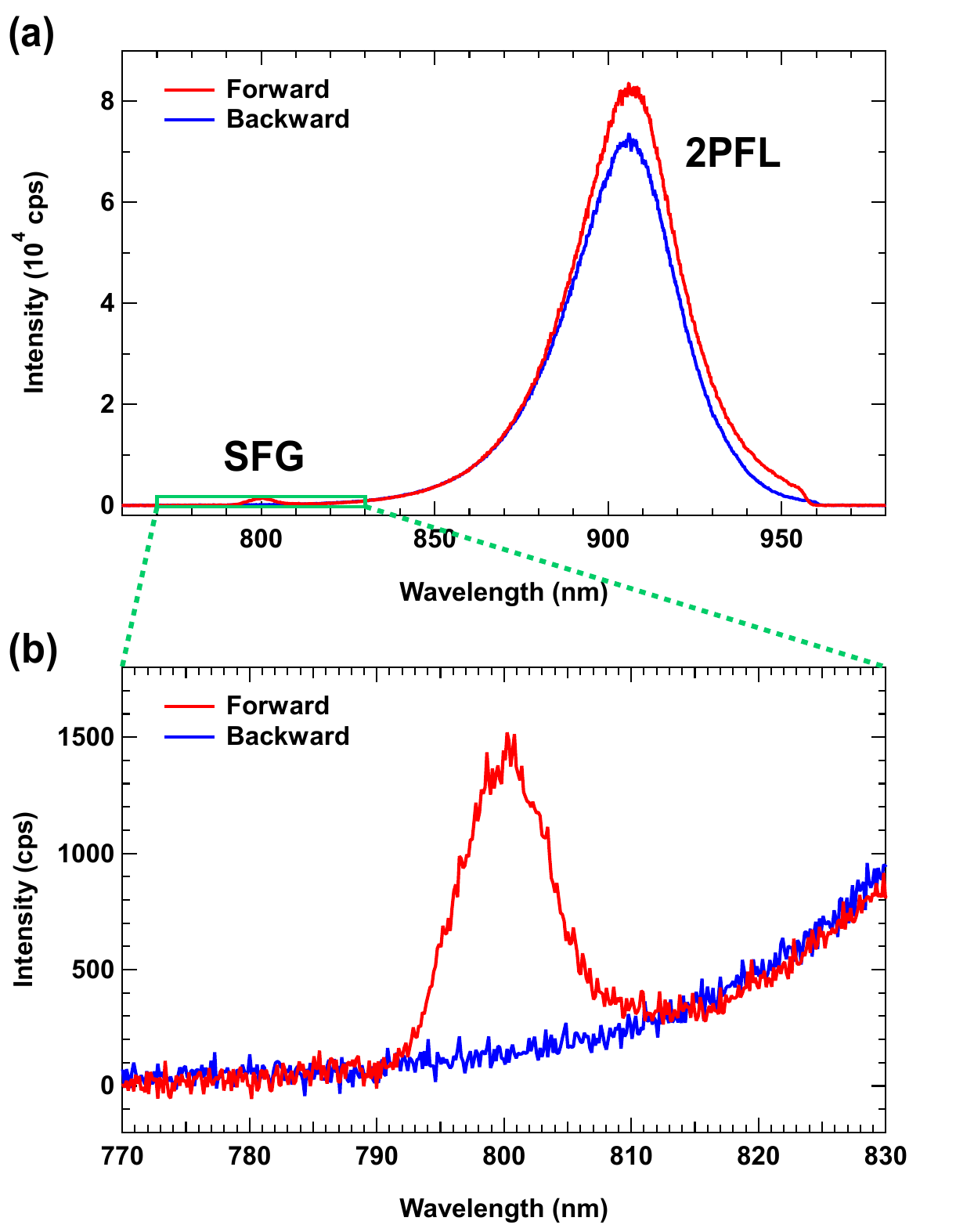}
 \caption{SFG and 2PFL spectra of GaAs.
 The detection was performed simultaneously in both forward (red) and backward (blue) directions.
 Panel (b) shows an enlarged view of the rectangular region indicated in panel (a).}
 \label{fig14}
\end{figure}

As described in Sec.~\ref{temp_prof_vis_pulse}, we measured the SFG signals from GaAs in a far-field geometry.
The input powers of the IR and 1033 nm beams were 20 and 1 mW, respectively.
In this measurement, the detection was performed simultaneously in both forward and backward directions.
A representative result is shown in Fig.~\ref{fig14},
which corresponds to the time delay $\tau=0\,\mathrm{fs}$.
Not only was the SFG signal observed in the forward direction (Fig.~\ref{fig14}(b)),
but the 2PFL signal was also detected
in both forward and backward directions (Fig.~\ref{fig14}(a)).
Because the band gap of GaAs is 1.4 eV (890 nm),
we observed 2PFL signals induced by the 1033 nm pulse alone and by both the 1033 nm and IR pulses.
This result clearly demonstrates that SFG is a coherent process that generates a signal
in the forward direction satisfying the phase-matching condition,
whereas the 2PFL signal is emitted isotropically as an incoherent process.
It should be noted that we routinely measure the SFG and 2PFL signals to optimize the optical alignment
for the forward and backward directions and to adjust the temporal overlap between the IR and 1033 nm pulses.

\section{Comparison between envelope of asymmetric visible pulse and its cross-correlation with IR pulse\label{app_time_envelope}}

\begin{figure}[H]
 \includegraphics[width=8cm]{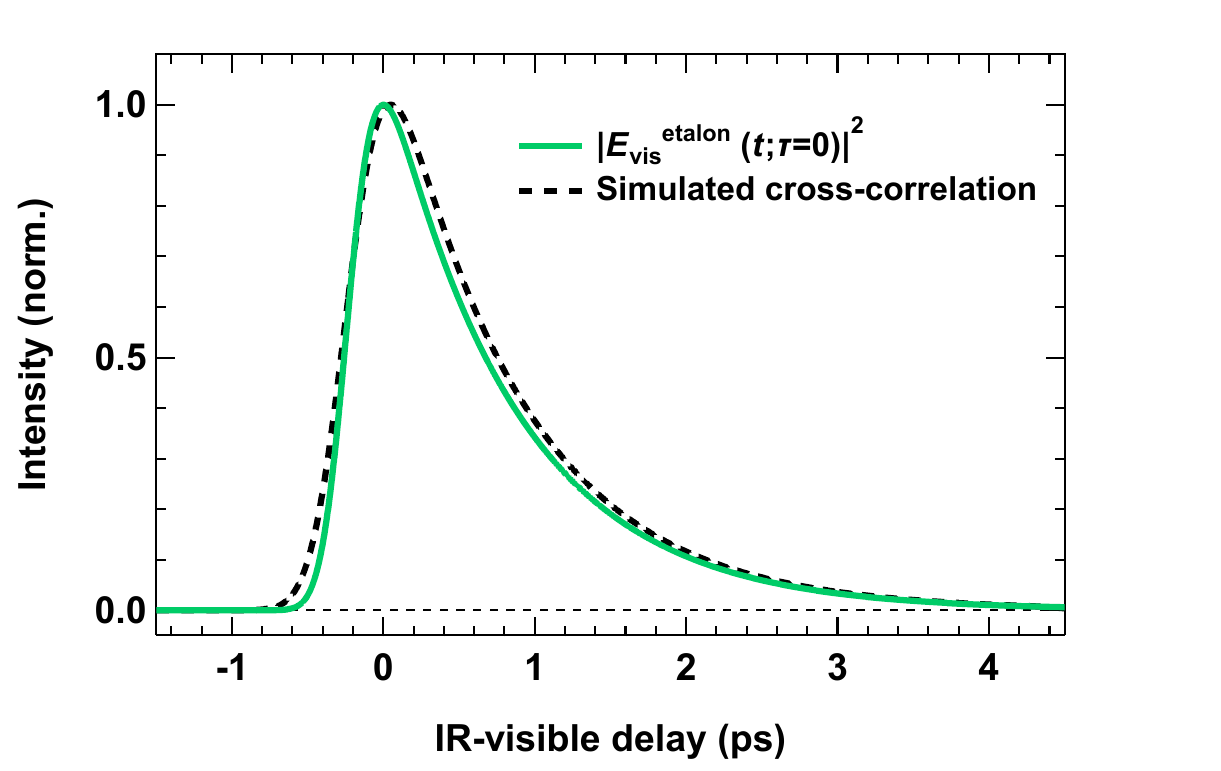}
 \caption{Calculated envelope of temporally asymmetric visible pulse
 generated by an air-spaced Fabry–P\'{e}rot etalon, $|E_\mathrm{vis}^\mathrm{etalon}(t)|^2$ (green curve),
 and simulated cross-correlation between temporally asymmetric visible and IR pulses (black dashed curve).
 Both curves exhibit almost identical shapes, which indicates that the observed cross-correlation in Fig.~\ref{fig8}
 directly reflects the characteristics of the temporally asymmetric visible pulse.}
 \label{fig15}
\end{figure}

\section{Comparison of STM images of the same scanned area obtained without and with laser irradiation\label{STM_images}}

\begin{figure}
 \includegraphics[width=6cm]{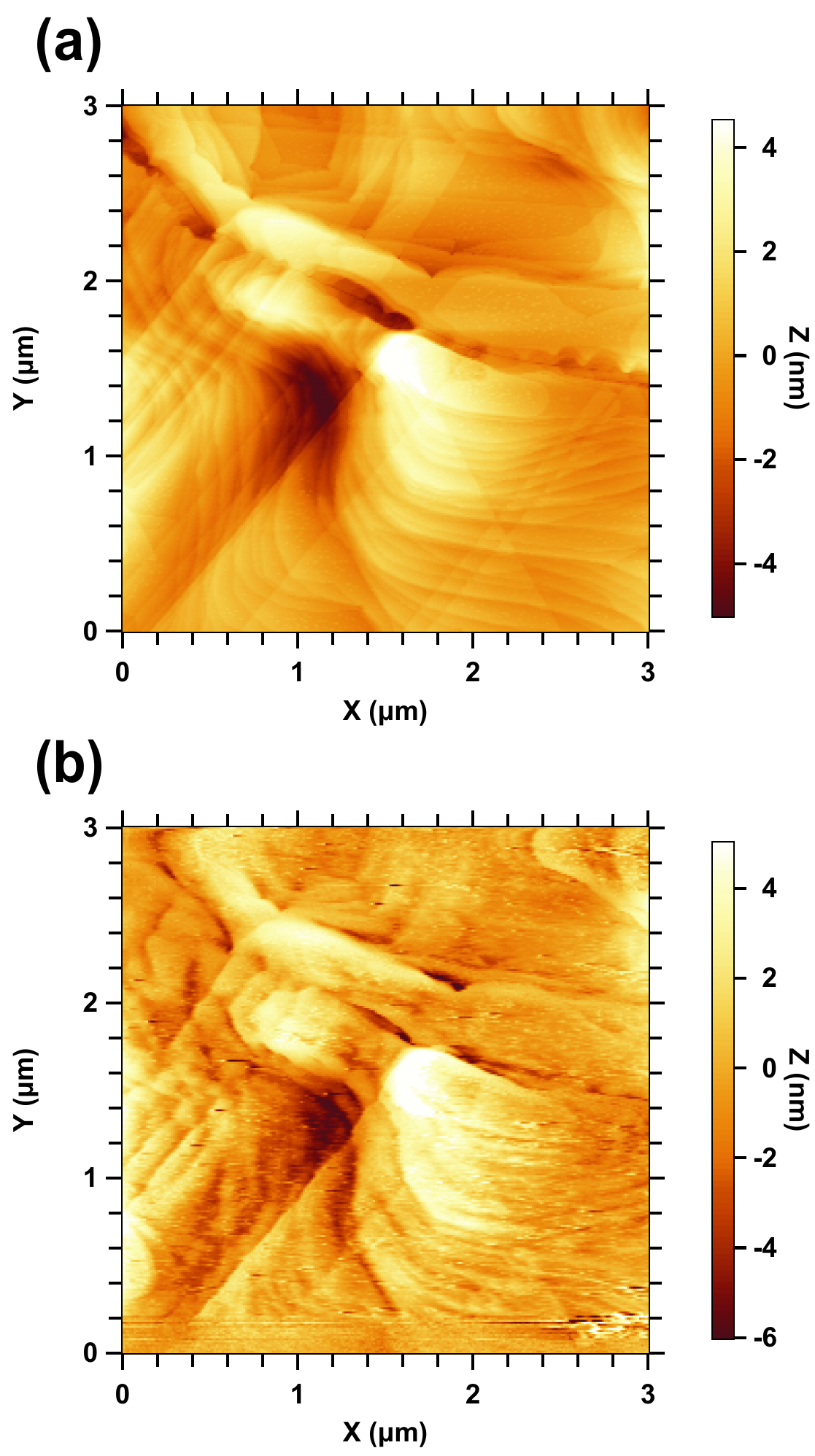}
 \caption{STM images of $3\,\mu\mathrm{m}\times3\,\mu\mathrm{m}$ area (256 × 256 pixels)
 of a 4-MBT SAM on Au(111).
 Color scale bars indicate height.
 (a) 0.5 nA, 0.1 V, without laser irradiation;
 (b) 2 nA, 0.05 V, with laser irradiation.
 The image shown in (b) was acquired during SFG measurement at 3000 cm$^{-1}$ (light blue curve in Fig.~\ref{fig9}(a)).}
 \label{fig16}
\end{figure}

Fig.~\ref{fig16} shows STM images of the same scanned area of a SAM of 4-MBT on Au(111),
acquired (a) without and (b) with laser irradiation.
Although the STM image became slightly blurred under laser irradiation,
the Au(111) steps remained visible, which indicated that nanoscale structures could still be resolved.
This blurring is attributed to fluctuations in the tunneling current,
induced by photon-assisted tunneling due to the enhanced electric field
within the tip–substrate plasmonic nanogap,
as discussed in a previous study.\cite{Sakurai_NL25}
It should be noted that the STM image and spectrum remained unchanged after a series of measurements,
suggesting that the sample was not damaged.

\section{Fitting parameters for curves in Fig.~\ref{fig11}\label{app_fitting_parameter}(a)}

The fitting parameters for Eq.~(\ref{I_SFG3}), corresponding to the curves in Fig.~\ref{fig11}(a), are listed in Table~\ref{tab1}.
Panels (a–d) correspond to IR pulse central wavenumbers of 2850 cm$^{-1}$ (orange),
2900 cm$^{-1}$ (yellow), 2950 cm$^{-1}$ (green), and 3000 cm$^{-1}$ (light blue), respectively.

\begin{table}[H]
 \caption{Fitting parameters for Eq.~(\ref{I_SFG3}), corresponding to the curves in Fig.~\ref{fig11}(a) at $\tau=0\,\mathrm{fs}$.}
 \begin{ruledtabular}
  \begin{tabular}{ccc}
   & (a) & (b) \\
   \hline
   $A_\mathrm{NR}$                           &  2.97   $\pm$ 0.02  & 3.96   $\pm$ 0.03 \\
   $\theta_\mathrm{NR}\,(^\circ)$      &  71.2   $\pm$ 5.8   &  51.8  $\pm$ 1.9  \\
   $A_q$                                     &  -4.47  $\pm$ 0.54  & -20.39 $\pm$ 1.03 \\
   $\omega_q\, (\mathrm{cm}^{-1})$           &  2861.2 $\pm$ 0.9   & 2909.7 $\pm$ 0.4  \\
   $\Gamma_q\, (\mathrm{cm}^{-1})$           &  10.7   $\pm$ 0.9   & 12.8   $\pm$ 0.4  \\
   $\omega_\mathrm{IR}^0\, (\mathrm{cm}^{-1})$           &  2849.1 $\pm$ 0.2   & 2895.1 $\pm$ 0.2  \\
   $\sigma_\mathrm{IR}\, (\mathrm{cm}^{-1})$ &  40.0   $\pm$ 0.3   & 38.5   $\pm$ 0.2
  \end{tabular}
 \end{ruledtabular}
 \vspace{5mm}
 \begin{ruledtabular}
  \begin{tabular}{ccc}
   & (c) & (d) \\
   \hline
   $A_\mathrm{NR}$                           &  3.79   $\pm$ 0.05  &  4.19   $\pm$ 0.04 \\
   $\theta_\mathrm{NR}\,(^\circ)$      &  27.3   $\pm$ 4.2   &  122.9  $\pm$ 7.3  \\
   $A_q$                                     &  -28.88 $\pm$ 2.61  &  -7.20  $\pm$ 0.87 \\
   $\omega_q\, (\mathrm{cm}^{-1})$           &  2924.6 $\pm$ 0.8   &  3008.0 $\pm$ 1.1  \\
   $\Gamma_q\, (\mathrm{cm}^{-1})$           &  15.9   $\pm$ 0.9   &  11.6   $\pm$ 1.1  \\
   $\omega_\mathrm{IR}^0\, (\mathrm{cm}^{-1})$           &  2952.8 $\pm$ 0.6   &  3002.5 $\pm$ 0.3  \\
   $\sigma_\mathrm{IR}\, (\mathrm{cm}^{-1})$ &  46.4   $\pm$ 0.6   &  44.5   $\pm$ 0.4
  \end{tabular}
 \end{ruledtabular}
 \label{tab1}
\end{table}

\section{Signal enhancement mechanisms: coherent SFG \textit{vs.} incoherent Raman processes\label{app_SFG_vs_Raman}}

\begin{figure}
 \includegraphics[width=6cm]{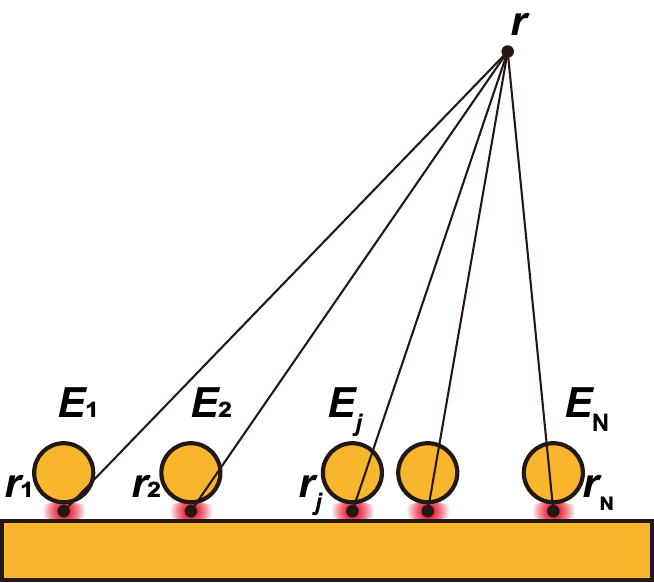}
 \caption{Schematic showing the contributions of individual electric fields
 at position $\bm{r}_j$ to the signal observed at position $\bm{r}$.}
 \label{fig17}
\end{figure}

The contrast between strong and weak enhancement mechanisms originates from the coherent nature of the SFG process.
We consider a system in which multiple metal particles are deposited on a metal substrate,
as schematically illustrated in Fig.~\ref{fig17}.
A locally enhanced near field is generated at each hot spot formed in the particle–substrate junction.
The electric field at position $\bm{r}$, originating from a hot spot located at $\bm{r}_j$, can be expressed as
\begin{equation}
 \bm{E}_j(\bm{r},t)
  = \tilde{\bm{E}}_j(\omega) e^{i[\bm{k}\cdot(\bm{r}-\bm{r}_j) - \omega t]}
  = \tilde{\bm{E}}_j(\omega) e^{i[\bm{k}\cdot(\bm{r}-\bm{r}_0) - \omega t + \theta_j]},
\end{equation}
where $\bm{E}_j(\bm{r},t)$ is the electric field observed at position $\bm{r}$ and time $t$,
$\tilde{\bm{E}}_j(\omega)$ is its amplitude, $\bm{k}$ is the wave vector, and $\omega$ is the angular frequency;
moreover, $\bm{r}_0=1/N\sum_{j=1}^N\bm{r}_j$ and $\theta_j=\bm{k}\cdot(\bm{r}_0-\bm{r}_j)$.
Then, the total electric field is given by the superposition of the individual contributions:
\begin{equation}
 \bm{E}(\bm{r},t)
  = \sum_{j=1}^N \bm{E}_j(\bm{r},t)
  = e^{i[\bm{k}\cdot(\bm{r}-\bm{r}_0) - \omega t]} \sum_{j=1}^N \tilde{\bm{E}}_j(\omega)e^{i\theta_j}.
\end{equation}
The corresponding signal intensity is
\begin{equation}
 I(\bm{r}) \propto |\bm{E}(\bm{r},t)|^2
  = \left|\sum_{j=1}^N \tilde{\bm{E}}_j(\omega)e^{i\theta_j} \right|^2.  \label{I_total}
\end{equation}
If the phases in Eq.~(\ref{I_total}) are randomly distributed,
destructive interference dominates, leading to weak enhancement.
In contrast, in highly ordered plasmonic structures, the fields interfere constructively,
resulting in strong enhancement.
In Raman scattering, which is intrinsically an incoherent process, Eq.~(\ref{I_total}) is replaced by
\begin{equation}
 I(\bm{r})
  \propto \sum_{j=1}^N \left|\tilde{\bm{E}}_j(\omega) \right|^2,
\end{equation}
indicating that the signals are simply additive and do not interfere.

\end{document}